\PassOptionsToPackage{unicode}{hyperref}
\PassOptionsToPackage{hyphens}{url}
\PassOptionsToPackage{dvipsnames,svgnames,x11names}{xcolor}
\documentclass[
  letterpaper,
  DIV=11,
  numbers=noendperiod]{scrartcl}

\usepackage{amsmath,amssymb}
\usepackage{setspace}
\usepackage{iftex}
\ifPDFTeX
  \usepackage[T1]{fontenc}
  \usepackage[utf8]{inputenc}
  \usepackage{textcomp} 
\else 
  \usepackage{unicode-math}
  \defaultfontfeatures{Scale=MatchLowercase}
  \defaultfontfeatures[\rmfamily]{Ligatures=TeX,Scale=1}
\fi
\usepackage{lmodern}
\ifPDFTeX\else  
\fi
\IfFileExists{upquote.sty}{\usepackage{upquote}}{}
\IfFileExists{microtype.sty}{
  \usepackage[]{microtype}
  \UseMicrotypeSet[protrusion]{basicmath} 
}{}
\makeatletter
\@ifundefined{KOMAClassName}{
  \IfFileExists{parskip.sty}{%
    \usepackage{parskip}
  }{
    \setlength{\parindent}{0pt}
    \setlength{\parskip}{6pt plus 2pt minus 1pt}}
}{
  \KOMAoptions{parskip=half}}
\makeatother
\usepackage{xcolor}
\makeatletter
\ifx\paragraph\undefined\else
  \let\oldparagraph\paragraph
  \renewcommand{\paragraph}{
    \@ifstar
      \xxxParagraphStar
      \xxxParagraphNoStar
  }
  \newcommand{\xxxParagraphStar}[1]{\oldparagraph*{#1}\mbox{}}
  \newcommand{\xxxParagraphNoStar}[1]{\oldparagraph{#1}\mbox{}}
\fi
\ifx\subparagraph\undefined\else
  \let\oldsubparagraph\subparagraph
  \renewcommand{\subparagraph}{
    \@ifstar
      \xxxSubParagraphStar
      \xxxSubParagraphNoStar
  }
  \newcommand{\xxxSubParagraphStar}[1]{\oldsubparagraph*{#1}\mbox{}}
  \newcommand{\xxxSubParagraphNoStar}[1]{\oldsubparagraph{#1}\mbox{}}
\fi
\makeatother

\providecommand{\tightlist}{%
  \setlength{\itemsep}{0pt}\setlength{\parskip}{0pt}}\usepackage{longtable,booktabs,array}
\usepackage{calc} 
\usepackage{etoolbox}
\makeatletter
\patchcmd\longtable{\par}{\if@noskipsec\mbox{}\fi\par}{}{}
\makeatother
\IfFileExists{footnotehyper.sty}{\usepackage{footnotehyper}}{\usepackage{footnote}}
\makesavenoteenv{longtable}
\usepackage{graphicx}
\makeatletter
\newsavebox\pandoc@box
\newcommand*\pandocbounded[1]{
  \sbox\pandoc@box{#1}%
  \Gscale@div\@tempa{\textheight}{\dimexpr\ht\pandoc@box+\dp\pandoc@box\relax}%
  \Gscale@div\@tempb{\linewidth}{\wd\pandoc@box}%
  \ifdim\@tempb\p@<\@tempa\p@\let\@tempa\@tempb\fi
  \ifdim\@tempa\p@<\p@\scalebox{\@tempa}{\usebox\pandoc@box}%
  \else\usebox{\pandoc@box}%
  \fi%
}
\def\fps@figure{htbp}
\makeatother
\NewDocumentCommand\citeproctext{}{}

\makeatletter
 \let\@cite@ofmt\@firstofone
 \def\@biblabel#1{}
 \def\@cite#1#2{{#1\if@tempswa , #2\fi}}
\makeatother
\newlength{\cslhangindent}
\newlength{\csllabelwidth}
\newenvironment{CSLReferences}[2] 
 {\begin{list}{}{%
  \setlength{\itemindent}{0pt}
  \setlength{\leftmargin}{0pt}
  \setlength{\parsep}{0pt}
  \ifodd #1
   \setlength{\leftmargin}{\cslhangindent}
   \setlength{\itemindent}{-1\cslhangindent}
  \fi
  \setlength{\itemsep}{#2\baselineskip}}}
 {\end{list}}
\usepackage{calc}

\usepackage{float}
\usepackage{tabularray}
\usepackage[normalem]{ulem}
\usepackage{graphicx}
\usepackage{rotating}
\UseTblrLibrary{booktabs}
\UseTblrLibrary{siunitx}
\NewTableCommand{\tinytableDefineColor}[3]{\definecolor{#1}{#2}{#3}}

\SetTblrStyle{foot}{font=\footnotesize}
\usepackage{pdflscape}
\KOMAoption{captions}{tableheading,figureheading}
\makeatletter
\@ifpackageloaded{caption}{}{\usepackage{caption}}
\AtBeginDocument{%
\ifdefined\contentsname
  \renewcommand*\contentsname{Table of contents}
\else
  \newcommand\contentsname{Table of contents}
\fi
\ifdefined\listfigurename
  \renewcommand*\listfigurename{List of Figures}
\else
  \newcommand\listfigurename{List of Figures}
\fi
\ifdefined\listtablename
  \renewcommand*\listtablename{List of Tables}
\else
  \newcommand\listtablename{List of Tables}
\fi
\ifdefined\figurename
  \renewcommand*\figurename{Figure}
\else
  \newcommand\figurename{Figure}
\fi
\ifdefined\tablename
  \renewcommand*\tablename{Table}
\else
  \newcommand\tablename{Table}
\fi
}
\@ifpackageloaded{float}{}{\usepackage{float}}
\floatstyle{ruled}
\@ifundefined{c@chapter}{\newfloat{codelisting}{h}{lop}}{\newfloat{codelisting}{h}{lop}[chapter]}
\floatname{codelisting}{Listing}

\makeatother
\makeatletter
\@ifpackageloaded{caption}{}{\usepackage{caption}}
\@ifpackageloaded{subcaption}{}{\usepackage{subcaption}}
\makeatother

\usepackage{bookmark}

\IfFileExists{xurl.sty}{\usepackage{xurl}}{} 
\hypersetup{
  pdftitle={Fifty Shades of Greenwashing: The Political Economy of Climate Change Advertising on Social Media},
  pdfauthor={, and },
  colorlinks=true,
  linkcolor={blue},
  filecolor={Maroon},
  citecolor={Blue},
  urlcolor={Blue},
  pdfcreator={LaTeX via pandoc}}

\title{Fifty Shades of Greenwashing: The Political Economy of Climate
Change Advertising on Social Media}
\author{Robert Kubinec\textsuperscript{$\dagger{}$,1} \and Aseem
Mahajan\textsuperscript{$\dagger{}$,2,*}}
\date{November 18, 2025}

\begin{document}
\maketitle
\begin{abstract}
In this paper, we provide a novel measure for greenwashing -- i.e.,
climate-related misinformation -- that shows how polluting companies can
use social media advertising related to climate change to redirect
criticism. To do so, we identify greenwashing content in 11 million
social-political ads in Meta's Ad Targeting Datset with a measurement
technique that combines large language models, human coders, and
advances in Bayesian item response theory. We show that what is called
greenwashing has diverse actors and components, but we also identify a
very pernicious form, which we call political greenwashing, that appears
to be promoted by fossil fuel companies and related interest groups.
Based on ad targeting data, we show that much of this advertising
happens via organizations with undisclosed links to the fossil fuel
industry. Furthermore, we show that greenwashing ad content is being
micro-targeted at left-leaning communities with fossil fuel assets,
though we also find comparatively little evidence of ad targeting aimed
at influencing public opinion at the national level. The supplementary
information can be downloaded from
\href{https://www.icloud.com/iclouddrive/00eqjqGpFLQ86sPSGGWRPuuhw\#mahajan\%5Fkubinec\%5FSI}{this
link.}\footnote{We thank Jieun Byeon, Ryan Lohr, Ellis Padgett, and
  Aliza Danish for invaluable research assistance. We thank New York
  University Abu Dhabi High Performance Computing Centre and the Texas
  A\&M ACES supercomputer for computing support. We thank attendees at
  the 2024 Language Models in Social Science conference for helpful
  comments. We acknowledge support from the National Science Foundation
  ACCESS program for computational resources.}
\end{abstract}

\setstretch{1.5}
\textsuperscript{$\dagger{}$}
These authors contributed equally to this work.

\textsuperscript{1} University of South Carolina\\
\textsuperscript{2} KAPSARC School of Public Policy

\textsuperscript{*} Correspondence:
\href{mailto:aseem.mahajan@kspp.edu.sa}{Aseem Mahajan
\textless{}aseem.mahajan@kspp.edu.sa\textgreater{}}

\newpage

\section{Introduction}\label{introduction}

Growing awareness of climate change over the last thirty years has
spurred political efforts to replace fossil fuels with renewable energy
through regulations, tax credits, and subsidies.\footnote{Data from the
  Climate Policy Radar Database shows nearly a twenty-fold increase in
  climate laws and policies between 1997 and 2018, from 72 to 1,500
  (Nachmany and Setzer 2018).} It has also shaped individual
preferences, giving rise to ``green consumers,'' who consider the
environmental impacts of companies or products in their purchasing
decisions (Peattie 2010). Between growing pressure from policymakers and
consumers, companies have increasing incentives to persuade the broader
public that their products help to alleviate the harms of warming to
avoid adverse regulatory decisions or consumer boycotts.

These trends raise concerns that firms may mislead the public about the
environmental effects of their practices or products, a practice
commonly described as \emph{greenwashing}.\footnote{The definition of
  \emph{greenwashing} presented here is a slight modification of one
  developed by Greenpeace, who describe it as ``the act of misleading
  consumers regarding the environmental practices of a company or the
  environmental benefits of a product or service.''} In recent years,
political agencies across continents have taken aim at corporate
greenwashing (Bryan 2023; Speed 2024; Ring and Speed 2023), levying
millions of dollars in fines.\footnote{Deutsche Bank, for instance, paid
  the SEC \$25 million for overstating the weight placed on
  environmental, social, and governance (ESG) factors in funds.} Over
2023 and 2024, United Kingdom (UK) regulatory authorities banned asset
managers from making misleading claims about sustainability (Bryan 2023)
and conducted ``wide-ranging clampdowns'' on consumer-facing companies
that make false environmental claims (Speed 2024; Ring and Speed 2023).
In the same period, the European Commission investigated twenty airlines
for making misleading greenwashing claims (European Commission 2024),
and the Australian Securities and Investment Commission (ASIC) took
civil action against Vanguard Investments for environmental, social, and
governance (ESG) claims (ASIC 2024).

In the context of climate change, firms that greenwash may overstate the
extent to which their corporate social initiatives regulate greenhouse
gas (GHG) emissions or counteract their effects.\footnote{An analysis of
  ExxonMobil's ``Energy Solutions'' advertisements by Plec and Pettenger
  (2012, 466) demonstrates how the campaign constructs a greenwashed
  frame ``by highlighting and amplifying science, technology, and the
  expertise of authorities'' to ``persuade consumers that oil companies
  will be the caretakers of our environment.''} Such behavior is in
keeping with a large body of research on political lobbying, which
demonstrates how private interests strategically appeal to the public to
achieve political ends (Kollman 1998; Baumgartner et al. 2009; Dür and
Mateo 2016; Dür 2019; McKnight and Hobbs 2013) and use voluntary
environmental self-regulation to preempt public regulation (Potoski and
Prakash 2011, 2004; Malhotra, Monin, and Tomz 2019).\footnote{The
  literature on self-regulation generally describes interactions between
  firms, who voluntarily choose to disclose potentially incriminating
  information about themselves, and regulators, who choose whether
  self-regulation is preferable to public regulation and, if so, how
  such a system should be designed. However, firms' incentives to
  self-disclosed remain the same even in a less organized environment.
  By self-disclosing to the public, firms may preempt them from
  demanding more stringent regulation from policymakers.} Firms hold
incentives to make misleading claims\footnote{As Dür (2019) points out,
  David Truman, in an early book about the interaction between interest
  groups and the public, writes that ``almost invariably one of the
  first results of the formal organization of an interest group is its
  embarking upon a program of propaganda, though rarely so labelled,
  designed to affect opinions concerning the interests and claims of the
  new group'' (Truman, David B 1951, 213).} --- indeed, the fossil fuel
industry has a long-history of doing so (Oreskes and Conway 2011, 6;
Supran and Oreskes 2017) --- and these persuasion efforts can have
political consequences.

We propose a theory of greenwashing that distinguishes between
greenwashing aimed at avoiding adverse regulatory action versus
advertising which is aimed at attracting climate-conscious consumers. We
argue that what we term \emph{political greenwashing} is more relevant
to policy outcomes concerning climate change. What we call
\emph{consumer greenwashing,} on the other hand, may meet some technical
definitions of greenwashing defined in the literature but is
theoretically distinct from the greenwashing aimed at achieving
corporate political aims. It is this former kind of greenwashing that we
believe to be of greater significance to policy efforts to control
climate change as its aim is to undermine costly efforts to move away
from fossil fuels.

To understand how political greenwashing occurs, we derive a novel
measure of greeenwashing from approximately 1 million Facebook and
Instagram advertisements in Meta's Ad Library and Ad Targeting Dataset
(collectively, ``Meta Ad Data''). Due to the complexity of the concept,
we employ methodological triangulation, including human coding, large
language models, and regular expression matching, which we then feed
into a robust Bayesian item response theory model (Kubinec 2025). From
this system, we create nuanced ad-level political greenwashing scores
with uncertainty and robustness to non-ignorable missingness.

As we show with these measures, the firms that are most affected by
political regulation are also those most likely to engage in political
greenwashing. Furthermore, political greenwashers appear to target
audiences where a fossil fuel company's assets are located. In these
environments, we show that advertisements with political greenwashing
will target influential demographic groups and areas where fossil fuel
companies face possible adverse regulatory decisions. At the same time,
we show that this type of advertising does not appear to be aimed at
influencing national-level public opinion. Social media advertising
seems most useful to businesses as a micro-level political strategy that
can protect important investments.

\section{Theory}\label{theory}

In this paper we are focusing on the role of economic actors with a
vested interest in undermining climate change policies, specifically
companies in the fossil fuel industry (Mildenberger 2020; Stokes 2020;
Lucas 2021; Tienhaara 2018). For these companies, a transition away from
fossil fuels would reduce the value of their fixed assets in production;
these cannot be easily converted to renewable production, with some
notable exceptions such as the growing geothermal industry (Anderson and
Rezaie 2019). Ownership of oil wells, natural gas pipelines, offshore
platforms and other assets is a rent that will lose value if governments
force consumers to transition to renewable energy. For these reasons,
these companies have a significant financial interest in opposing
regulation that would reduce demand for fossil fuel companies up to the
value of these rents. We should expect based on standard political
economy theory that these firms should be willing to undertake costly
actions to protect these investments, and indeed there is a substantial
empirical record of such actions (Mulvey et al. 2015).

At the same time, this view of the firm can be considered reductionist
in that it ignores the possibility that shareholders and managers could
exercise a form of enlightened self-interest in which they support an
energy transition while they deplete their existing fossil fuel assets
(Nasiritousi 2017). The theory of \emph{corporate social responsibility}
proposes that companies can indeed undertake relatively altruistic
actions, especially when there are related benefits to the firm through
improving its image to consumers (Carroll, Primo, and Richter 2016).
Fossil fuel companies are actively involved in CSR (Megura and Gunderson
2022), and in principle they could support climate change alleviation
efforts out of an awareness of their responsibility for the damage to
the environment caused by their fossil fuel production. Fossil fuel
involvement in CSR could also be strategic to avoid blowback from
consumers or policymakers who might see any obstructionist political
activity as behavior that should be punished (Bergman 2018).

For these reasons, there are competing incentives that make the
strategic use of advertising on social media by fossil fuel companies an
open question. Will fossil fuel companies adopt a relatively altrustic
strategy that supports climate change policies out of a sense of
corporate social responsibility, or will they seek to persuade voters
that climate change is less of a problem and governments should avoid
adopting costly climate change policies? The individual nature of social
media, which permits individual fossil fuel companies to create their
own accounts and target audiences at the micro level (Macdonald,
Gunderson, and Widner 2024), creates an opening for this type of
behavior to be observed with much more precision than was previously
possible. Furthermore, social media offers corporations the ability to
advertise with plausible deniability because they can run ads through
accounts affiliated with third parties that are ostensibly independent
entities.

This research question is of broader significance as well because fossil
fuel companies are some of the central actors in the growing climate
crisis. Social media permits these companies the chance to influence
public opinion in a way that is unprecedented in human history, and we
as yet lack a clear understanding of how this type of political activity
works and how effective it might be. For these reasons, we theorize that
this type of activity could depend as well on the type of political
institutions in a given country. Where public opinion is quite
important, which would generally happen in regimes that are more
accountable and responsive (Slater 2008), then we might expect that
companies will have a greater incentive to strategically shape public
opinion on climate change in an anti-regulation direction. In less
democratic countries, by contrast, fossil fuel companies might need to
rely more on strong relationships with elites as institutions are by
definition not as responsive to the median voter (Kakenmaster 2024;
Francis and Kubinec 2025).

To study this research questions, we will need to also build a
conceptual foundation for the behavior we want to study: so-called
greenwashing by corporations. If corporations are using social media to
influence public opinion in anti-regulatory direction, then they are
likely engaging in greenwashing by diverting attention from costly
policies towards solutions that are more amenable to continued fossil
fuel production. In other words, it is unlikely that fossil fuel
companies will simply argue that global warming does not exist or that
citizens should ignore it; rather, they will employ more sophisticated
outreach strategies that confuse and re-direct climate concern. For
these reasons, we focus on greenwashing as a viable advertising strategy
that could undermine climate action in a way that is difficult to
detect, permitting fossil fuel companies to achieve their strategic
objectives while minimizing blowback.

Because research on greenwashing spans disciplines and measures
different behaviors, papers on the subject likewise use various data and
definitions to assess public claims made by actors. In a systematic
review of the ``greenwashing literature,'' Lyon and Montgomery (2015)
finds that most quantitative empirical papers use companies' annual
reports (Wiseman 1982; Ingram and Frazier 1980), corporate social
responsibility (CSR) reports/disclosures to investors or nonprofit
organizations\footnote{One frequently source of data are company
  disclosures to the Carbon Disclosure Project nonprofit organization
  that assists public and private actors disclose information about
  their environmental impact.} (Mahoney et al. 2013; Uyar, Karaman, and
Kilic 2020; Tang and Demeritt 2018) and their websites to assess claims
about environmental performance. Similar sources of data used in
existing literature include corporate indices such as Bloomberg ESG
scores (Yu, Van Luu, and Chen 2020; Yu, Guo, and Luu 2018; Tamimi and
Sebastianelli 2017) and lists of corporate environmental scores
published by news publications (Du et al. 2014).

Besides these studies, which examine greenwashing across various
industries, others measure greenwashing using ecolabels (Sirieix et al.
2013) or ISO 14001 certification, earned by meeting standards in
voluntary environmental audits (Potoski and Prakash 2005; Blackman
2012). These studies use firms as their unit of analysis. Additionally,
public statements or signals of environmental compliance in these
studies are directed to a narrow group of stakeholders, such as
investors or regulators.

While greenwashing can be done via any medium, the rise of social media
has given corporations and interest groups a novel platform which which
to shape their image, as existing research has shown. Using Twitter's
academic API, Blazkova et al. (2023) examine half a million tweets
containing the keywords `greenwash', `greenwashing', and `greenwashed,'
performing a content analysis and identifying key stakeholders involved
in these threads. Similarly, Amin, Ali, and Mohamed (2024) analyze
approximately 170,000 tweets issued from the accounts of companies on
the FTSE 350 Index, using a structural topic model to measure the
adoption and extent of corporate social responsibility (CSR) disclosure.
Finally, Kwon et al. (2024) analyzes Instagram posts between 2019 and
2021, focusing on posts issued by the 15 most reputable companies in the
world, as measured by GlobalRepTrak 100, a ranking of corporate
reputation issued by RepTrak, a reputation intelligence platform.
Analyzing nearly 17,000 posts, it identifies 394 green advertising
posts.

Although there is a substantial amount of research on greenwashing,
there is little agreement in the literature about how the concept is
defined or measured. Given that the aim of this paper is in
understanding greenwashing by fossil fuel companies, we propose that
greenwashing should be differentiated between \emph{political}
greenwashing and \emph{consumer} greenwashing. We argue that these two
types of greenwashing have different aims, and while they both may
include information about climate change and corporate behavior that is
misleading in some sense, they are unlikely to have similar effects in
terms of shaping public opinion. Climate change regulatory policy will
affect fossil fuel producers much more heavily than most companies, and
for that reason, many companies would prefer to bolster their climate
change bona fides out of an effort to win over climate-conscious
consumers rather than in an attempt to reduce fears over climate change.

Separating these two types of greenwashing requires nuance and an
awareness of the context of the company behind the advertising. As we
explain later, our measurement strategy uses a diverse array of
indicators to differentiate types of greenwashing. Ultimately, we want
to isolate that type of greenwashing that could benefit fossil fuel
companies by diverting the public's concern over climate change. There
are many other types of climate change-related advertising, some of
which can fit definitions of greenwashing as proposed in the literature,
but we argue that greenwashing that is primarily aimed at attracting
climate-conscious consumers using claims that may be more or less
accurate is of less importance to the study of the political economy of
climate change. For these reasons, our main aim in this study is to see
to what extent fossil fuel producers and related actors are engaged in
political greenwashing, and more specifically, how they might be doing
so via social media accounts, whether via official or unofficial
accounts.

\subsection{Hypotheses}\label{hypotheses}

Once we have overcome the measurement challenge, our research design
incorporates relatively straightforward observational inference. Due to
our theoretical priors that we specified above, we think that higher
levels of political greenwashing among fossil fuel producers and their
allies would be evidence that these companies are attempting to
undermine costly policy efforts to oppose climate change. While our
research design is not causally identified, we still have reason to
believe that these associations are informative as we do not think there
are relevant confounders that could produce a spurious assocation. In
other words, if we can find associations between fossil fuel producers
and greenwashing, we believe that this association is most likely due to
the companies' investment in fossil fuels rather than some other
unobserved factor. This confidence derives from the research we cited
earlier about the profit maximization incentive of corporations and the
real and present danger that costly climate change policies pose to
their bottom line.

For these reasons, we want to collect data on the different types of
advertisers who are engaging in any kind of climate change advertising.
These other actors comprise the null hypothesis or reference group by
which we can evaluate the possible greenwashing of fossil fuel
companies. Furthermore, we will gain important descriptive information
from having a more general sense of the state of climate change
advertising that will be of interest to policymakers and other scholarly
work on the topic.

We also want to collect information on the identity of a variety of
types of interest groups with potential links to fossil fuel companies.
Part of the strategic nature of corporate advertising is that it may
come from relatively neutral-seeming trade associations or think tanks.
If corporations are funding these interest groups, then the climate
change advertising that these other organization undertake is
essentially directed by the funders but with the deniability of a
separate organizational platform. For these reasons, we do not restrict
ourselves only to fossil fuel companies and their subsidiaries but also
interest groups with which they share some relationship.

With this combined dataset of the identity of advertising actors and the
level of greenwashing, we aim to test the following hypothesis:

\begin{quote}
H1: Fossil fuel companies and related interest groups are more likely
than other climate change advertisers to engage in political
greenwashing.
\end{quote}

In addition to this hypothesis, we want to further understand to what
extent this type of advertising is related to political institutions and
to the nature of political competitions. We hypothesize that if our
theory is correct then we should be able to detect more greenwashing in
countries where popular legitimacy is more important; i.e., in
democratic countries. We further think that greenwashing should be more
evident among groups that are important for electoral reasons,
especially older demographics that are more likely to vote.

In the United States, members of the Baby Boomer generation hold the
largest share of wealth (Board of Governors of the Federal Reserve
System 2025), have the highest average net worth (Board of Governors of
the Federal Reserve System 2023) and vote at higher rates than other
generations (Pew Research Center 2023). Consequently, political and
economic actors may have greater incentives to influence their behavior
and misrepresent their actions. In contrast, individuals under 18 are
least likely to make household purchasing decisions or vote.
Greenwashing could therefore be less prevalent in this cohort.
Advertisers may also hold a longer-time horizon with regard to younger
generations, choosing to avoid greenwashing to avoid damaging their
brand.

To better understand the extent to which greenwashing is associated with
political institutions and demographics, we aim to test the following
hypothesis:

\begin{quote}
H2: Greenwashing ads are more likely to occur in more democratic
countries and among older demographic groups.
\end{quote}

An important feature of social media ads is micro-targeting, or the
ability to only show ads to very specific subsets of users based on
geographic and other criteria (Kreiss 2017; Dobber, Fathaigh, and
Borgesius 2019). Based on the theory we have proposed, we have reason to
believe that corporations would also micro-target ads, specifically if
there are existing assets or investment opportunities in the given area.
This micro-targeting approach contrasts with H3, which examines
prevalence of greenwashing across the social media landscape. As a
result, we want to test a third and final hypothesis:

\begin{quote}
H3: The presence of fossil fuel investments in a given geographic area
is associated with a greater prevalence of greenwashing content targeted
to that area.
\end{quote}

\section{Data}\label{sec-meta-dataset}

To obtain real-world measures of political and consumer-oriented
greenwashing, the data used in this paper conducts a comprehensive
analysis of the universe of \emph{all} Social Issues, Electoral, and
Political ads in English classified by Meta (Facebook's parent company),
which includes all ads related to environmental issues.\footnote{See
  section Section~\ref{sec-meta-dataset} for a description of Meta's
  classification scheme.} In addition to ads run by corporations, the
dataset includes statements by nongovernmental and international
organizations, political actors, and interest groups. As Lyon and
Montgomery observe, existing greenwashing literature ``shows that just
like corporations, NGOs, and governments can engage in greenwashing;
indeed, they may often serve as partners in corporate greenwashing''
(2015). The scope of actors included in the dataset allows the paper to
examine greenwashing conceived more broadly. The Meta Ad Datasets have
the added benefits of including specific data about the audience
targeted by advertisers. This allows the paper to present new
information about which audiences companies intend to address and which
audiences are exposed to greenwashing. Finally, advertisers must pay
money to run ads on Meta's platforms, and audiences may therefore pay
these ads more attention than social media posts from other individuals.

As we explain further below, the Ad Library includes information about
the delivery of advertisements while the Ad Targeting dataset describes
the audience that the advertiser seeks to target.

\subsection{Meta Ad Data}\label{sec-description-siep}

Facing criticisms about its advertising technology\footnote{Leading into
  and following the 2016 United States (US) presidential election,
  Facebook faced accusations of purveying misinformation and
  contributing to political polarization Levin (2017). In the same year,
  a ProPublica investigated revealed that Facebook's advertising
  technology allowed marketers to exclude specific ethnic groups by race
  (Angwin and Parris, Jr. 2016), leading to the US Department of Housing
  and Urban Development to lodge a formal complaint against the company
  (Isaac and Hsu 2021).}, Facebook revised its terms of service in 2018
to limit the scope of advertising allowed on its platform, regulate
advertisers, and increase transparency (Bouchaud 2024). In 2019,
Facebook, Inc.~(rebranded as Meta in 2021) introduced its Ad Library, a
collection of information about ads focusing on SIEP advertisements
(Shukla 2019). Beginning in May 2022, Meta began publishing this
information, along with ad targeting information, for SIEP ads run on
the platform starting from August 2020 (Meta Platforms 2024). Ads
classified as SIEP include (Meta Business Help Center 2025):

\begin{itemize}
\tightlist
\item
  Those made by a political actor or organization or advocates for an
  election outcome;
\item
  Ads about any election, referendum, or ballot;
\item
  Ads regulated in the advertiser's country as political advertising;
  and
\item
  Ads about any social issues in the place where the ad is being placed.
\end{itemize}

The final category include local sensitive or heavily debated topics,
particularly those that ``seek to influence public opinion through
discussion, or debate or advocacy for or against important topics, like
health and civil and social rights'' (Meta Business Help Center 2025).
For the entire European Union and fourteen countries, Meta has developed
a top-level country-specific list of topics considered sensitive. The
topic of environmental politics is included in all of the lists. To
identify and review SIEP ads, Meta relies on self-reporting by
advertisers; automated technology that reviews ad images, video, text,
targeting information, and landing pages; and manual reviews of ads. Ads
are reviewed on a monthly basis and sometimes reclassified.

The Meta Ad Library and Ad Targeting Dataset comprise the Meta Ad Data
used in this paper (Meta Platforms 2024). The Ad Library includes
information about the delivery of advertisements. This includes text of
the advertisement, the page associated with the advertisement, the
period over which the advertisement ran on Meta, and information about
the audience reached by the ad (e.g., demographic distribution, number
of impressions, etc.). The Ad Targeting Dataset, which maps to entries
in the Ad Library, describes the audience that an advertiser sought to
target with a particular ad. It provides information about the targeted
audience's location, age, gender, and language. Additionally, it
describes whether and how to target the ad to audience's connections
(Meta Platforms 2024). To ensure consistency throughout the analysis, we
focus on ads appearing in the Ad Library as of December 20, 2024.

\section{Measurement}\label{measurement}

To capture political greenwashing, as opposed to more anodyne forms of
grenwashing, we need to employ diverse types of indicators with respect
to climate change discourse, using both quantitative and qualitative
measurement strategies. By doing so, we can cast a broad net to measure
the complexities and nuances of political greenwashing in order to
overcome the ability of corporations and related interest groups to mask
such advertising. We further want to combine robust measures of
greenwashing with covariates that relate to the identity of the
advertising entity so that we can understand what types of actors
undertake greenwashing. In particular, we want to understand how much
greenwashing relates to political-economic theories of corporations with
strategic reasons to prefer less climate change regulation.

The measurement challenge we need to overcome is separating greenwashing
from less strategic forms of discourse around climate change. This task
is no trivial matter as it is easy for an actor to claim that it does
not want climate change to occur, i.e., cheap talk. The difference
between discourse that diverts attention from costly solutions to
climate change and a discourse that has an aim of mitigating climate
change can be one of considerable nuance. Furthermore, if our theory is
correct that fossil fuel producers are more likely to politically
greenwash in order to secure valuable political concessions, then these
companies will have both the incentive and resources to make their
advertising efforts relatively subtle. For these reasons, our
measurement strategy triangulates across both qualitative and
quantitative measurement techniques in order to best identify the latent
trait we want to capture.

\subsection{Qualitative Coding of Political
Greenwashing}\label{qualitative-coding-of-political-greenwashing}

Our qualitative approach to measuring greenwashing involves the use of
regular expressions to match specific keywords and employing human
coders to identify social media accounts with known linkages to the
fossil fuel industry.

\begin{longtable}[]{@{}lr@{}}
\caption{Number of observations by
keyword}\label{tbl-obs-count}\tabularnewline
\toprule\noalign{}
Description & Number of Observations \\
\midrule\noalign{}
\endfirsthead
\toprule\noalign{}
Description & Number of Observations \\
\midrule\noalign{}
\endhead
\bottomrule\noalign{}
\endlastfoot
\emph{carbon capture} & 34,547 \\
\emph{carbon remov}* & 1,194 \\
\emph{climat}* & 124,838 \\
\emph{coal} & 10,711 \\
\emph{extinction} & 6,848 \\
\emph{fossil fuel} & 11,974 \\
\emph{global warming} & 4,214 \\
\emph{greenhouse} & 7,195 \\
\emph{icecap} and \emph{melt} OR \emph{icecap} and \emph{flood} & 24 \\
\emph{recycle} & 22,825 \\
\emph{sustainable} & 71,605 \\
\textbf{any keywords} & 235,032 \\
\end{longtable}

Regular expression matching is used at two stages in our research
design. First, we employ a broad list of terms because we need to filter
out from Meta's massive Ad Library Dataset---which has over 10 million
ads---those ads that do not have any relationship to climate change (or
even the environment). At the same time, we need a list of terms that is
broad enough to capture all ads that have at least some positive
probability of being greenwashing. Beginning with basic words like
``climate change'', we developed a broad list of 191 terms using
embeddings, i.e., we provide a prompt to large language models to
provide synonyms of these key words. Given that Meta's database includes
ads from all over the world, we translate the keywords and regular
expressions into 12 commonly-spoken languages--Arabic, Bengali,
Urdu/Hindi, Tagalo, Mandarain, Indonesian, Portuguese, French, Spanish,
Swahili, Russian, and German--so that we can obtain the universe of
climate-related ads. The regular expressions we employed are shown in SI
section 3.

As a second filtering step, we removed ads that were primarily deployed
for electoral purposes, such as those referencing a specific candidate
or campaign. A large share of the SIEP ads are those run by U.S.
electoral campaigns, and electoral campaigns are outside the scope of
our study as they have a different aim than influencing the pro- or
anti-climate views of the population. This step is especially important
for identifying political greenwashing as we need ads that are about
climate change, have a politically-relevant theme, but are not appealing
for respondents to vote for a specific candidate.

The filtered list of ads included 1,176,612 ads from the larger dataset.
With this smaller, yet still quite large, list, we then coded ads for
the presence of specific terms which we had a prior reason to believe
could be related to greenwashing, including a mention of technologies
known to be used as excuses for climate action by fossil fuel companies,
including clean coal and carbon capture. The particular way we
operationalize these indicators using regular expressions is shown in
Table~\ref{tbl-obs-count}. Each ad had a value of 1 if it matched the
term and 0 otherwise, and these terms were also translated into other
languages for non-English ads.

For our second form of qualitative measurement, we hired human coders to
create a list of fossil fuel companies and related interest groups, and
then to comb through Meta's list of pages with related ads to see if
there were matches. To identify advertisements by fossil fuel companies,
we supplement a list of the top investor- and state-owned companies
ranked by potential emissions from reserves (Heede 2013, 2014; Heede and
Oreskes 2016) with a manually compiled list of fossil fuel companies,
interest groups, and affiliates. The interest groups were derived from
policy reports by the climate change think tank InfluenceMap\footnote{See
  https://influencemap.org/ for more information.} that identify
linkages between fossil fuel companies and other interest groups. In
total, we identify 18 pages associated with fossil fuel companies or
affiliates in the Meta Ad Data in English, which ran 2,155 ads. Of
these, we identified 240 ads by fossil fuel actors that were also
related to climate change.

In addition to InfluenceMap, we used existing published lists to
identify fossil fuel entities in the social media data. Using production
records from annual reports, public libraries, websites, and US Security
and Exchange Commission reports, and other sources, Heede (2014)
calculates historical fossil fuel production and carbon emissions for 90
entities producing more than eight million tonnes of carbon per year.
From this analysis, Heede and Oreskes (2016) focuses on seventy
companies and eight government-run entities producing 63\% of fossil
fuels between 1750 and 2010.\footnote{oreskes\_footnote} It calculates
potential emissions from the reserves of the seventy investor- and
state-owned companies, using self-reported reserves data. We conduct
searches for pages associated with these companies or companies with
similar names using regular expressions. After eliminating pages with
similar but unrelated names, this search produces sixty-nine pages
associated with twenty-one companies. Additionally, it produces seven
page names associated with five related interest groups or affiliates.
To add to this list, we (i) review the names of all page names in the
Meta Ad Data in English, identifying and investigating pages whose names
suggest that they might be associated with fossil fuel companies; and
(ii) performing a search for Meta pages associated with interest groups
in an external list of groups developed by InfluenceMap.

With this qualitative measurement, we are able to create further binary
variables for our ads dataset for whether a page is an oil company, a
related interest group, think tank, or other type of actor with some
known link to the fossil-fuel industry. In total we were able to
identify 233 pages in the Meta Ad Library that had such a known link.

\subsection{Quantitative Measurement
Evaluation}\label{quantitative-measurement-evaluation}

We also made use of quantitative techniques for measuing discourse given
the large scale of the data. However, it can be very difficult to
capture nuance from more simple text analysis models like the ``bag of
words'' approach; an ad could contain many references to carbon capture,
coal, and carbon removal yet still be aimed at increasing rather than
reducing the climate policy response. For these reasons, we introduce an
ensemble of open-source large language models (LLMs) as a way of
providing much more nuanced coding of greenwashing. An LLM is trained on
a large corpus of diverse texts and represents these texts in a
continuous latent space known as ``embeddings'' (Rodriguez, Spirling,
and Stewart 2023). Given an input text, LLMs will select the
highest-probability tokens conditional on the input, which can be almost
completely unstructured (Weber and Reichardt 2024).

To increase our measurement validity, we employ six open-source LLMs:
Mistral:7b (Jiang et al. 2023), Llama 3.2:3b (Touvron et al. 2023), Phi3
(Abdin et al. 2025), Gemma2 (Team et al. 2024), Deepseek-r1 (Guo et al.
2025), and Qwen2.5:7b (Yang et al. 2025). We use open-source models so
that we can better ensure reproducibility by setting seeds and storing
model weights (Barrie, Palmer, and Spirling 2024). These models are
small to medium LLMs that can be operated with a single standard 16-GB
GPU, which makes our method feasible without access to very large GPUs
or expensive computational resources.

The prompt we provide the LLMs is as follows:

\begin{quote}
Advertisement Text: {[}``ad text to classify''{]} : You are a human bot
evaluating a number of advertisements on webpages. You are given the
text of the advertisement and name of the website after \_\_\_\_\_.

Based on this information, your task is to answer the following
question:

Greenwashing is the act of making false or misleading statements about
the climate impact of a product or practice. It can be a way for
companies to maintain or increase their greenhouse gas emissions. Is
this ad engaged in greenwashing?

Your must choose exactly one of the following predefined answers: yes,
no

You will only respond with the answer. Do not repeat these instructions
or include the word `Answer:' before giving your answer. Do not give any
explanations or notes.
\end{quote}

While the text of the prompt constrained the LLM to an extent, the LLM
still sometimes returned additional unstructured text in its response.
As a result, we coded the response as a 1 if we could identify a
positive response in its answer, and 0 if a negative response, and in
cases where we could not identify an answer, we coded the response as
missing (i.e., a hallucination).

To directly evaluate the LLM, we randomly selected 1,000 ad texts and
assigned them to be coded as either having or not having greenwashing by
three research assistants (RAs). Our RAs had at least undergraduate or
graduate degrees in the social sciences, and were given the same prompt
as the LLM for the purposes of fairness. To facilitate a comparison with
LLMs and their extensive training data, we allowed the RAs up to five
minutes per coding to search for background information about the ad. We
report the results of this evaluation exercise in SI Section 1. In
short, the LLMs had a strong tendency to over-classify ads as containing
political greenwashing relative to RAs, and the Llama 3.2:3b model
showing the strongest overlap with human coding decisions. As such,
while it is clear from the exercise that LLMs can identify this type of
speech, they are also relatively noisy when considered individually.

Besides LLMs, we also include a model known as DEBATE (Burnham 2025)
that is a BERT formulation (Bidirectional Encoder Representations from
Transformers). This model has been trained to identify stances in
political texts. To employ the model, we have it determine whether a
specific hypothesis (i.e., declarative statement), namely whether the ad
in question opposes efforts to mitigate climate change, is supported for
a given ad. While this is a conceptually simpler prompt than the one we
give to LLMs, this type of information is quite useful for estimating
our greenwashing scores as we describe later. We code our ads with this
model with a 1 if they are pro-climate change action and 0 otherwise.

\subsection{Combining Measures Into a Greenwashing
Score}\label{combining-measures-into-a-greenwashing-score}

All of the methods of measurement we have mentioned previously have
limitations. The unstructured nature of LLMs, for example, can also
produce random sequences of text or a complete misreading of the prompt
(Tonmoy et al. 2024). Qualitative measures are difficult to replicate
for the entire dataset, and relatively simple measures like binary
indicators for words are unlikely to capture sufficient nuance. However,
each of these measurement strategies has its own strengths as well. For
these reasons, we want to find a way to incorporate as much of the
information in each individual measure while also safeguarding against
non-ignorable measurement error.

To produce an aggregate greenwashing score at the ad level we turn to
the ideal point model parameterized via Bayesian item-response theory
(Bafumi et al. 2005; Clinton, Jackman, and Rivers 2004; Kubinec 2025).
The advantage of using IRT to aggregate across diverse indicators is to
allow us to incorporate measurement error in a principled way. For
example, by using the LLM output as input to a measurement model, we can
accept that LLMs will sometimes fail or ``hallucinate'' and allow that
uncertainty to propagate as measurement error (Burnham 2023; Green et
al. 2025). A structured measurement model will also give valuable
insights into how these indicators are plausibly related to each other
and to the latent concept we seek to measure.

We do not further define IRT or Bayesian measurement models here as we
note there is a robust literature on the topic (Bafumi et al. 2005;
Clinton, Jackman, and Rivers 2004; Park 2011; Hanson and Sigman 2021;
Pemstein, Meserve, and Melton 2017). Our particular variant of IRT is
known in political science as the ideal point model given that it will
provide latent scores that can be interpreted as the ideal points of
legislators (Clinton, Jackman, and Rivers 2004). However, as we are
scaling ad texts rather than persons, we note that this variant of IRT
allows us to have bipolarity in the item indicators, that is, our
possible indicators for greenwashing can be either positively or
negatively related to greenwashing. The model estimates will provide us
with both the sign and magnitude of this relationship via the item
discrimination parameters.

However, the flexibility of the ideal point model comes with a price as
it requires informative priors on at least some parameters to identify
the rotation and scale of the latent variable (Bafumi et al. 2005). Our
IRT specification via Kubinec (2025) uses a re-scaled Beta distribution
for the item discrimination parameters as the primary means of
identification of the latent variable; this distribution forces all item
discrimination parameters to lie within the open interval of (-1,+1). To
identify the sign of the latent variable, we further assign a strong
positive prior on our indicators for the phrase ``natural gas'' of
\(\text{Beta}(1000,.001)\) and a strong negative prior on the phrase
``fossil fuel'' of \(\text{Beta}(.001,1000)\). We use these priors by
examining the relative frequency of our indicators for ads that we
manually classified as belonging to fossil fuel companies---we found
that ``natural gas'' appeared much more often in these ads and ``fossil
fuel'' appeared much less. As such, we believe that these priors will be
helpful to identify the latent trait of political greenwashing. In
addition, the stochastic nature of our measurement model does not
require a mechanistic relationship between these informative priors and
the latent trait. So long as these relationships hold on average in the
data (i.e., are more right than wrong), then we can identify a unique
rotation of our greenwashing scores that is conditional on this prior
information about the likely nature of greenwashing.

There is another important facet of the data that we need to
incorporate--the ability of LLMs to ``hallucinate'' or return responses
that are outside the scope of the prompt or otherwise irrelevant. To
permit us to incorporate this type of non-ignorable missigness in the
data along with parallelism to handle the large data size, we employ the
parameterization of Kubinec (2025) that explicitly accounts for
potential bias in missing data by fitting a separate first-stage IRT
model for whether the measurement is missing. By doing so, we obtain
missingness item discrimination parameters as described by Kubinec
(2025) to allow us to adjust our scores for LLMs potentially
hallucinating more often when ads have very high or very low levels of
greenwashing. We employ default priors for other parameters as specified
by Kubinec (2025).

We note that our empirical approach also breaks new ground in applying
emerging tools -- specifically, LLMs -- to difficult measurement
challenges. While LLMs have received enormous attention in industry and
academia in recent years (Manning, Zhu, and Horton 2024; Hackenburg and
Margetts 2024; Dubova, Moskvichev, and Zollman 2024; Marshall et al.
2024), it can be difficult to incorporate them into robust research
designs because LLM output is not always reliable due to
``hallucinations'' (Tonmoy et al. 2024), or in more conventional terms,
non-classical measurement error. For these reasons, our research design
discussed in this section is aimed at taking advantage of the remarkable
breadth of LLMs while adding additional information from external
sources to anchor estimates via IRT.

\section{Results}\label{results}

\subsection{Greenwashing Scores}\label{greenwashing-scores}

\begin{figure}

\caption{\label{fig-greenscores}Distribution of Posterior Uncertainty
Intervals for Ad-level Greenwashing Scores}

\centering{

\pandocbounded{\includegraphics[keepaspectratio]{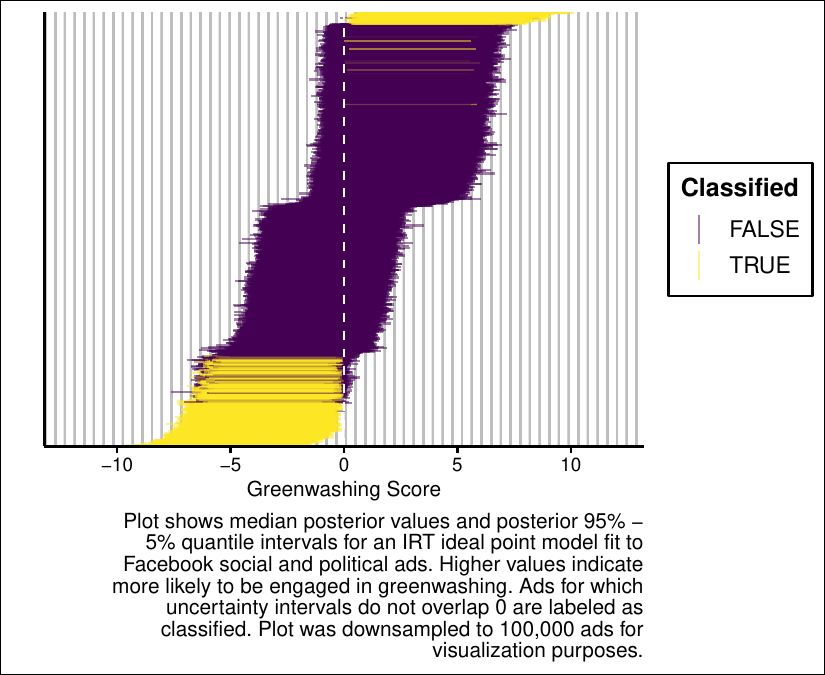}}

}

\end{figure}

We next turn to a discussion of the greenwashing scores that were
derived from the measurement model that we described earlier. Because
these results derive from a principled Bayesian item-response theory
model, we have uncertainty in the scores at the ad level even with the
limited number of observations per ad. We show the full distribution of
the mean of the posterior draws (i.e., the point estimates) with 5\% to
95\% quantile uncertainty intervals in Figure~\ref{fig-greenscores}.
Positive values of the score indicate a higher probability of
greenwashing while lower values indicate a lower probability of
greenwashing, though the exact value of the score is based on an
arbitrary scale centered around zero as is convention in IRT estimates.

What is immediately evident from Figure~\ref{fig-greenscores} is that
there is substantial uncertainty in the scores at the ad level. This
uncertainty derives from the fact that we have 15 indicators per ad.
While this is a reasonable number for an index, we have only observation
of these indicators per ad. For this reason, while the spread of the
mean scores is quite large, the model can only classify approximately
21.8\% of ads as being either related or unrelated to greenwashing with
high probability, with 3.1\% of ads classified with political
greenwashing and 18.7\% without greenwashing. While the number of
positively-classified ads with high probability is relatively small, we
note that this number is reasonably close to the proportion of
greenwashing we estimated with human coders at 1.6\% (95\% CI, 0.9\% -
2.5\%) (see SI Section 1).

\begin{table}

\caption{\label{tbl-top10}Top Ten Ad-level Greenwashing Scores}

\centering{

\centering
\begin{tblr}[         
]                     
{                     
width={1\linewidth},
colspec={X[0.636363636363636]X[0.0909090909090909]X[0.0909090909090909]X[0.0909090909090909]X[0.0909090909090909]},
column{2,3,4,5}={}{font=\fontsize{0.8em}{1.1em}\selectfont, halign=c,},
column{1}={}{font=\fontsize{0.8em}{1.1em}\selectfont, halign=l,},
}                     
\toprule
Ad Text & Language & 5\% Low Score & Posterior Mean & 95\% High Score \\ \midrule 
++ The motto must be less Qatar, not more! ++ Human rights violations, a strict ban on homosexuality, exploitation of migrant workers, corruption in the World Cup award: For Minister-President Stephan Weil (SPD), Qatar is nevertheless a reliable & German & 3.06 & 7.05 & 10.9 \\
New Mexico is a leader in energy production through oil and natural gas. We can't allow the far Left to keep the state from producing. & English & 3.06 & 6.99 & 11.7 \\
‪Happy to announce \$16.2M to build a 92km natural gas pipeline from Peace River to La Crete with MLA Dan Williams.   ‪This strategic infrastructure project will provide safe and secure supply for residents and increase the potential to grow the reg & English & 3.06 & 6.71 & 11.5 \\
Oil and natural gas is keeping our economy strong! Stand up to radical environmentalists to ensure we have a vibrant Texas economy. & English & 2.65 & 6.55 & 10.7 \\
+++ Russia Day in Mecklenburg-Vorpommern: Good international relations are a guarantee for prosperity in Europe +++ Today, Russia Day took place for the fourth time in Mecklenburg-Vorpommern. The chairman of the AfD parliamentary group, @706006922 & German & 2.57 & 6.53 & 10.6 \\
+++Farle calls for criminal sanctions in the Bundestag+++Criminal elements: intent, deception, and fraud proven+++Budget debate on November 28, 2023+++ While Scholz, in his government statement, declares business as usual in his technocratic lang & German & 2.11 & 6.33 & 10.6 \\
Federal government must temporarily use even more coal to survive the winter despite gas shortages. The fears of the BVB/FREIE WÄHLER parliamentary group from the spring have come true: The federal government has been failing to conserve natural ga & German & 1.97 & 6.27 & 10.5 \\
In response to Gov. Kevin Stitt's call to eliminate the state income tax, Senate President Pro Tempore Greg Treat listed a potential GPT increase on oil and natural gas as one way to offset the decrease in state tax revenue.  Even though Oklahoma p & English & 2.05 & 6.22 & 10.2 \\
\#Opinion \#Fedumop \#TransportationSector \#Massification \#GNV |🔴 The constant change of ministers in the Transport portfolio delays the agreements that the unions reach with this ministry. The president of the Federation of Multimodal Transport of Pi & Spanish & 2.38 & 6.07 & 10 \\
Join a group of oil and natural gas lovin' Texans! & English & 2.19 & 6.07 & 9.9 \\
\bottomrule
\end{tblr}

}

\end{table}%

We validate the model in two different ways. First, we look at the ads
that were classified as having the highest probability of greenwashing,
which we show in Table~\ref{tbl-top10}. Ads in other languages are
translated into English, and for space limitations, we only show the
first 250 characters of the ads (some of the ads only mention climate
change related issues later in the post). The German ads are calling for
international cooperation for or against gas exporting countries (i.e.,
Qatar and Russia), while the English ads are more focused on supporting
oil and gas industries in particular U.S. states. We note that the
posterior uncertainty of these scores is quite high, but even so, the
vast majority of these ads would fit the core ideas of greenwashing that
we discussed previously, albeit with contextual features that are unique
to a given social and political context.

\begin{figure}

\caption{\label{fig-items}Posterior Distributions for Item
Discrimination Parameters from Greenwashing IRT Model}

\centering{

\pandocbounded{\includegraphics[keepaspectratio]{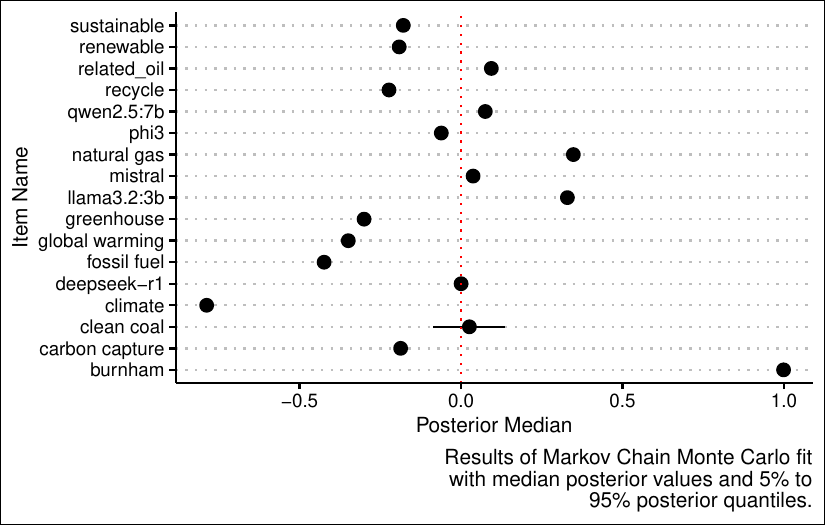}}

}

\end{figure}%

Second, we can examine the item discrimination parameters from the IRT
model in Figure~\ref{fig-items}. A discrimination parameter is very
useful for this measurement exercise because it shows which indicators
(items) ``load'', or have the highest information, about the latent
trait we are trying to measure. It is important to note that the
discrimination parameters have additive effects, so even smaller
discrimination values can matter for measuring the latent trait given
that they still contain some unique information. On the other hand, a
discrimination parameter of zero would indicate no information about the
latent trait--the indicator has essentially dropped out of the model.

Figure~\ref{fig-items} shows the posterior mean value of the
discrimination parameters for each indicator with the 5\% to 95\%
posterior quantile to capture uncertainty. Uncertainty for these scores
is quite small given that the number of items (15) is small relative to
the number of ads, resulting in a large amount of data per parameter. As
discussed in the methods section, we added informative prior information
that the presence of the term ``natural gas'' should positively predict
political greenwashing while the presence of the term ``fossil fuel''
should negatively predict political greenwashing. We do indeed see these
indicators on the assigned sides, though they are not the strongest
factors. Rather, the strongest indicator for our political greenwashing
score is our binary classification from the DEBATE BERT model about an
ad being pro- or anti-policy action towards climate change.

\begin{figure}

\caption{\label{fig-items-miss}Posterior Distributions for Item
Missingness Discrimination Parameters from Greenwashing IRT Model}

\centering{

\pandocbounded{\includegraphics[keepaspectratio]{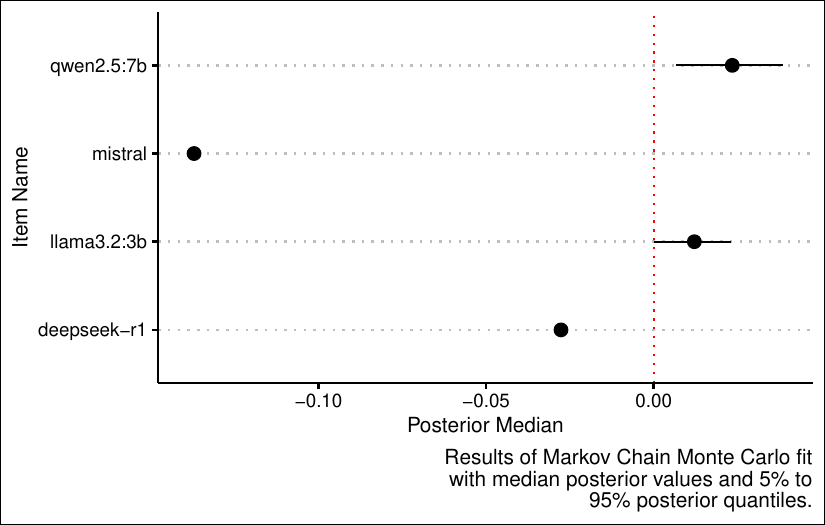}}

}

\end{figure}%

For the LLMs, we find in Figure~\ref{fig-items} that Llama3.2 is the
second strongest positive predictor of greenwashing (the natural gas
indicator is technically the second-highest but it had a strongly
positive informative prior). The other LLMs have noticeably less
discriminatory power for the greenwashing score. Most of the other
important indicators are terms, such as ``greenhouse'' and ``global
warming'' having a strong relationship with non-greenwashing ads. As
such, in our measurement model we have more indicators that are
primarily helpful in determining which ads \emph{do not} contain
greenwashing than we do in confirming greenwashing. This likely explains
why in Figure~\ref{fig-greenscores} we have more ads strongly classified
as non-greenwashing than greenwashing.

What we also want to note is that the words ``sustainable'' and
``recycling'' are negative predictors of political greenwashing. This
relationship provides evidence in support of our contention that there
are fundamentally two types of greenwashing; our model is primed to
capture \emph{political} greenwashing while words like ``sustainable''
and ``recycling'' tend to be associated either with either
\emph{consumer} greenwashing or content that is authentically advocating
for action to prevent climate change. There could be content in this ads
that is not strictly true from the perspective of climate science, but
at the same time the discourse is not aimed at obfuscating the climate
change movement.

In addition, in Figure~\ref{fig-items-miss} we can see the missingness
discrimination parameters for the LLM models (the other indicators do
not have them as they did not have any missing data). As mentioned
earlier, these missingness parameters are estimated in a first-stage
model to endogenize missing data. These missingness parameters show what
proportion of information about the latent trait of political
greenwashing is present in the missingness patterns introduced by LLM
hallucination. Interestingly, while LLama3.2 is the most informative
model on observed data, Mistral's \emph{missing} data is in fact fairly
informative at predicting lower greenwashing scores. In other words,
given the prompt, Mistral tended to hallucinate more when presented with
a non-greenwashing ad. The 2-stage IRT model was than able to backwards
infer that when Mistral hallucinated, the ad should have a lower score.
Through this method, we can safeguard inferences from LLM model
breakdown.

\subsection{Exploratory Analysis of Ad
Targeting}\label{exploratory-analysis-of-ad-targeting}

Before we test hypotheses, we want to describe the general patterns in
this large and diverse collection of data. While not strictly related to
our hypotheses, this section points to important features of the ad
targeting data that are relevant to current concerns about the nature of
climate change advertising and its links to the fossil fuel industry.

\subsubsection{Fossil Fuel Industry
Networks}\label{fossil-fuel-industry-networks}

An issue of significant concern is that the relatively unregulated
nature of social media advertising would permit fossil fuel industry
groups to run ads that are not directly connected with their official
accounts. Our universe of social media ads provides a unique window into
this type of behavior as we can look for similarities in patterns of ad
text across plausibly unrelated groups. To do so, we employed an
embedding model to classify the Facebook Pages in terms of their ad text
cosine similarity (Sanh et al. 2020). We consider any two Pages to be
connected if they have at least five ads with a greenwashing score 1 SD
above the average and a cosine similarity score of at least +0.8 (the
maximum is +1). In other words, the pages must have been running at
least five very similar ads that had a reasonably high probability of
greenwashing content.

\begin{figure}

\caption{\label{fig-networks}Networks of High Greenwashing Ad Similarity
in Ad Library}

\centering{

\pandocbounded{\includegraphics[keepaspectratio]{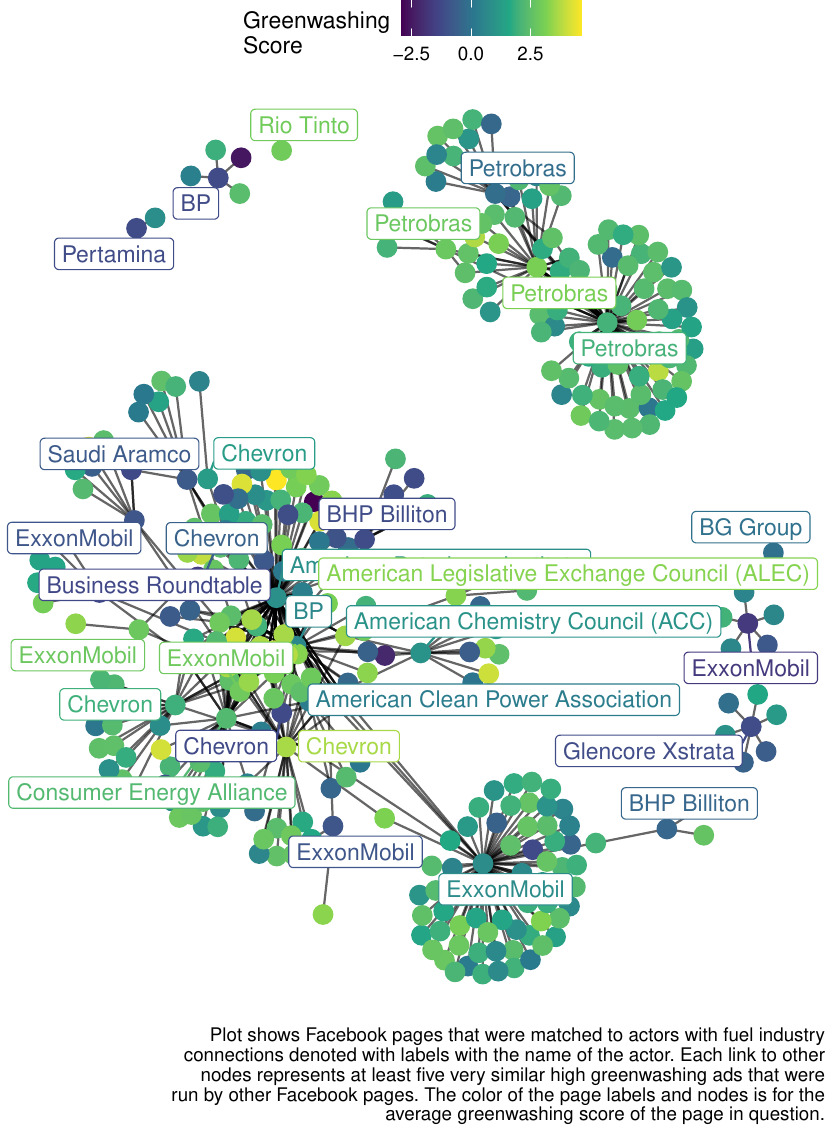}}

}

\end{figure}

To narrow down the list to those pages that might have links to the
fossil fuel industry, we filter the resulting connected pages to only
those pages that were pre-identified via external qualitative
information as having a known identity as a fossil fuel actor (whether
corporation or interest group). This step results in a final network of
31 pages that have plausible greenwashing-related links to 352 other
pages that we did not pre-identify as fossil fuel actors. This network
of connected pages is shown in Figure~\ref{fig-networks} in which the
identified pages of well-known interest groups and corporations are
labeled. We further colored the nodes by the average greenwashing scores
for all ads for each page. As can be seen, there is a very large and
dense network around U.S. and U.K. fossil fuel actors, particularly the
American Petroleum Institute, Chevron, and BP. There is also a distinct
cluster around Petrobras, the Brazilian state oil producer.

It is important to note that these links do not necessarily mean that
there is an actual relationship or that there is any undisclosed
advertising funding. Indeed, a closer inspection of the connected nodes
shows that a considerable number of the pages report their connection to
the fossil fuel actor, such as distinct ExxonMobil subsidiaries like
ExxonMobil México. However, it is also clear that these associations are
not always reported whether via the Facebook Page name or via the
funding entity data reported when a political ad is run on Facebook. For
example, the American Chemistry Council, which sometimes advocates on
behalf of the fossil fuel industry,\footnote{See
  \url{https://www.motherjones.com/politics/2025/01/fossil-fuel-industry-talking-points-obstruction-climate-action-delay-denial/}.}
has greenwashing ad content similarities with Enbridge Gas, the Gas
Networks of Ireland, the Propane Education and Research Council, the
Petroleum Alliance of Oklahoma, and the Hoover Institution. While
adjudicating the exact nature of these links would require extensive
qualitative investigation, the analysis does reveal notably similar
patterns of climate change discourse across seemingly disparate
organizations.

In Table~\ref{tbl-avg-diff}, we show the 10 pre-identified Facebook
pages with the biggest positive difference in greenwashing scores with
the linked page. A high positive difference indicates that the
pre-identified page's ads have much less greenwashing content on average
than the linked page. While this statistic does not provide proof of a
subversive relationship, it does indicate that these pages have notable
similarities in ads while still maintaining on the whole very different
average profiles. As a result, they may be able to mask discursive
similarities between different pages that would very hard for users to
discern. One notable example is the second-highest difference, which is
between BHP Billiton, an Australian mining company with oil and gas
assets, and NZ Energy Voices, a New Zealand NGO that ``tells the story
of natural gas and its importance to New Zealand''\footnote{See
  https://x.com/nzenergyvoices} and derives its funding from the
Petroleum Exploration and Production Association of New
Zealand.\footnote{See
  https://www.energyresources.org.nz/news/giving-a-voice-to-the-people-new-campaign-launched-on-oil-and-gas-exploration-halt/}
While it is not clear if BHP Billiton currently has active oil and gas
operations in New Zealand, or simply promotes similar types of ads to NZ
Energy Voices, the obvious overlap in potential interests suggests that
there may indeed be networks (whether formal or informal) that
coordinate advertising on climate change-related issues.

\begin{table}

\caption{\label{tbl-avg-diff}Linked Facebook Pages with Greatest
Difference in Average Greenwashing Scores}

\centering{

\centering
\begin{talltblr}[         
entry=none,label=none,
note{}={Table shows Facebook pages with pre-identified fossil fuel links and linked pages where a link equals at least five very similar greenwashing ads. The selected pages were those with the greatest difference in greenwashing scores between the pre-identified page and the connected page, which could imply unacknowledged advertising.},
]                     
{                     
width={0.7\linewidth},
colspec={X[]X[]X[]X[]},
row{1}={}{font=\fontsize{1em}{1.3em}\selectfont,},
cell{2}{4}={}{font=\fontsize{1em}{1.3em}\selectfont, halign=c,},
cell{3}{4}={}{font=\fontsize{1em}{1.3em}\selectfont, halign=c,},
cell{4}{4}={}{font=\fontsize{1em}{1.3em}\selectfont, halign=c,},
cell{5}{4}={}{font=\fontsize{1em}{1.3em}\selectfont, halign=c,},
cell{6}{4}={}{font=\fontsize{1em}{1.3em}\selectfont, halign=c,},
cell{7}{4}={}{font=\fontsize{1em}{1.3em}\selectfont, halign=c,},
cell{8}{4}={}{font=\fontsize{1em}{1.3em}\selectfont, halign=c,},
cell{9}{4}={}{font=\fontsize{1em}{1.3em}\selectfont, halign=c,},
cell{10}{4}={}{font=\fontsize{1em}{1.3em}\selectfont, halign=c,},
cell{11}{4}={}{font=\fontsize{1em}{1.3em}\selectfont, halign=c,},
cell{2}{1}={}{font=\fontsize{1em}{1.3em}\selectfont, valign=m,},
cell{2}{2}={}{font=\fontsize{1em}{1.3em}\selectfont, valign=m,},
cell{2}{3}={}{font=\fontsize{1em}{1.3em}\selectfont, valign=m,},
cell{3}{1}={}{font=\fontsize{1em}{1.3em}\selectfont, valign=m,},
cell{3}{2}={}{font=\fontsize{1em}{1.3em}\selectfont, valign=m,},
cell{3}{3}={}{font=\fontsize{1em}{1.3em}\selectfont, valign=m,},
cell{4}{1}={}{font=\fontsize{1em}{1.3em}\selectfont, valign=m,},
cell{4}{2}={}{font=\fontsize{1em}{1.3em}\selectfont, valign=m,},
cell{4}{3}={}{font=\fontsize{1em}{1.3em}\selectfont, valign=m,},
cell{5}{1}={}{font=\fontsize{1em}{1.3em}\selectfont, valign=m,},
cell{5}{2}={}{font=\fontsize{1em}{1.3em}\selectfont, valign=m,},
cell{5}{3}={}{font=\fontsize{1em}{1.3em}\selectfont, valign=m,},
cell{6}{1}={}{font=\fontsize{1em}{1.3em}\selectfont, valign=m,},
cell{6}{2}={}{font=\fontsize{1em}{1.3em}\selectfont, valign=m,},
cell{6}{3}={}{font=\fontsize{1em}{1.3em}\selectfont, valign=m,},
cell{7}{1}={}{font=\fontsize{1em}{1.3em}\selectfont, valign=m,},
cell{7}{2}={}{font=\fontsize{1em}{1.3em}\selectfont, valign=m,},
cell{7}{3}={}{font=\fontsize{1em}{1.3em}\selectfont, valign=m,},
cell{8}{1}={}{font=\fontsize{1em}{1.3em}\selectfont, valign=m,},
cell{8}{2}={}{font=\fontsize{1em}{1.3em}\selectfont, valign=m,},
cell{8}{3}={}{font=\fontsize{1em}{1.3em}\selectfont, valign=m,},
cell{9}{1}={}{font=\fontsize{1em}{1.3em}\selectfont, valign=m,},
cell{9}{2}={}{font=\fontsize{1em}{1.3em}\selectfont, valign=m,},
cell{9}{3}={}{font=\fontsize{1em}{1.3em}\selectfont, valign=m,},
cell{10}{1}={}{font=\fontsize{1em}{1.3em}\selectfont, valign=m,},
cell{10}{2}={}{font=\fontsize{1em}{1.3em}\selectfont, valign=m,},
cell{10}{3}={}{font=\fontsize{1em}{1.3em}\selectfont, valign=m,},
cell{11}{1}={}{font=\fontsize{1em}{1.3em}\selectfont, valign=m,},
cell{11}{2}={}{font=\fontsize{1em}{1.3em}\selectfont, valign=m,},
cell{11}{3}={}{font=\fontsize{1em}{1.3em}\selectfont, valign=m,},
}                     
\toprule
Fossil Fuel Actor & Linked Page & Linked Page Funder & Greenwashing Difference \\ \midrule 
ExxonMobil & Atlantic Gulf  Pacific Company & NA & 5.2 \\
Chevron & Your Energy & American Gas Association & 4.8 \\
BHP Billiton & NZ Energy Voices & NZ Energy Voices & 4.7 \\
ExxonMobil & New Mexico Oil  Gas & New Mexico Oil and Gas Association & 4.4 \\
ExxonMobil & Grow Louisiana Coalition & Grow Louisiana Coalition & 4.2 \\
Chevron & Power In Cooperation & ConocoPhillips Company & 4.2 \\
ExxonMobil & Além da Superfície & INSTITUTO BRASILEIRO DE PETROLEO E GAS & 3.7 \\
BP & BOLD PAC & CHC BOLD PAC & 3.6 \\
Business Roundtable & ExxonMobil & EXXON MOBIL CORPORATION & 3.5 \\
BP & Texas Oil and Gas Association & Texas Oil and Gas Association & 3.5 \\
\bottomrule
\end{talltblr}

}

\end{table}%

It is important to note too that this exercise is necessarily
conservative as we limited our network data to those pages with
pre-identified fossil fuel associations. One page that was in the
broader list (i.e., the list of all pages with greenwashing connections
to other pages) but did not have any previous association with the
fossil fuel industry is the Empowerment Alliance, which as of late 2025
had 24,000 followers on Facebook.\footnote{See
  \url{https://www.facebook.com/TheEmpowermentAlliance/}.} The page
describes its mission as ``fighting for millions of Americans who care
about keeping energy affordable, reliable, and clean.''\footnote{See
  https://www.facebook.com/TheEmpowermentAlliance/.} While this group's
stated purpose might make it sound like a pro-climate change mitigation
platform, investigations by The Energy \& Policy Institute have revealed
a 501(c)4 ``dark money'' organization behind the page that is ``aligned
with the gas industry'' and has a president who is a former lobbyist for
Chesapeake Energy. In our data, the average greenwashing score for
Empowerment is +4.55, but it has no disclosed links to fossil fuel
institutions that we were able to identify with pre-existing
sources.\footnote{See
  https://energyandpolicy.org/the-empowerment-alliance-2/ for more
  information.}

\subsection{Inference About
Greenwashing}\label{inference-about-greenwashing}

\subsubsection{Hypothesis 1: Fossil Fuel Companies and
Greenwashing}\label{hypothesis-1-fossil-fuel-companies-and-greenwashing}

\begin{figure}

\caption{\label{fig-typegreen}Distribution of Posterior Mean
Greenwashing Scores for Ads by Entity Type}

\centering{

\pandocbounded{\includegraphics[keepaspectratio]{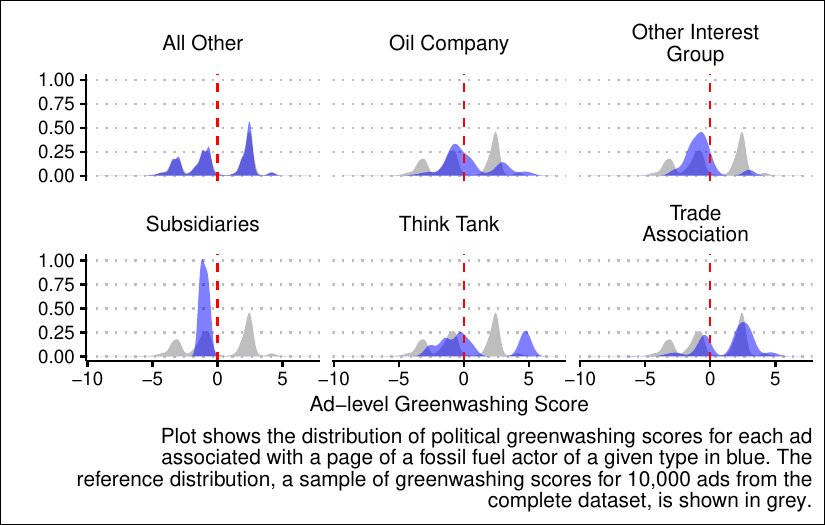}}

}

\end{figure}

We now turn to an evaluation of the hypotheses that we proposed at the
beginning of the article. We test our hypothesis 1 by examining
distributions of ads to provide a nuanced understanding of the patterns
in the data rather than relying on a single summary measure like a
\(p\)-value. To do so, we report the complete distribution of the
posterior mean greenwashing scores for the different types of entities
that we coded (fossil fuel companies, their subsidiaries, think tanks,
trade associations, and related interest groups) as blue distributions
in Figure~\ref{fig-typegreen}. In this plot, we show each type of fossil
fuel-related entity as a separate facet with the greenwashing scores of
the rest of climate ads as a reference distribution in gray.

On the whole, do these entities have higher greenwashing scores? The
answer largely depends on the entity type. One type of entity, oil
company subsidiaries, do not appear to have very much greenwashing
content in their ads. In general, the score distributions are also very
multi-modal, suggesting that these advertising platforms design some ads
with greenwashing content but also that they put out ads that are much
less easy to classify as greenwashing. The entity types with the highest
modes at positive greenwashing scores are the ``think tank'' category.
Given that we derived our list of interest groups from research on
entities with links to fossil fuel companies, we find this empirical
relationship to be very interesting as it is suggestive that fossil fuel
companies may be using unconventional platforms to mask greenwashing
advertising on social media. Furthermore, it is important to note that
there is considerable heterogeneity within the category of oil companies
as it includes both national state-owned oil companies like Petrobras
and private oil companies like Shell and BP.

The advantage of visually examining the distributions rather than
relying on summary statistics is that we can examine features that a
comparison of means would obscure (i.e., multi-modal distributions). It
is clear from Figure~\ref{fig-typegreen} that while there is reason to
believe that there is greenwashing occurring among these groups that we
pre-identified as having an association with the oil industry, the most
pronounced modes are seen only in the think tank category. Furthermore,
this behavior does not manifest itself in oil company subsidiaries,
probably because the subsidaries are engaged in lower-level retail
marketing rather than political action.

On the whole, we find moderate support for H1, with the primary caveat
that the relationship appears to vary significantly across the differen
types fossil fuel-related groups.

\subsubsection{Hypothesis 2: Country-level
Analysis}\label{hypothesis-2-country-level-analysis}

In order to test hypotheses 2 and 3, we employ a metric that enables
fair comparisons given different sizes in advertising markets. To that
end, we use in this section impression-weighted greenwashing scores by
different geographical areas and demographic groups. In other words, for
a given group, we multiply the ad greenwashing scores times the share of
impressions that each ad had for that target market. As such, the models
we employ in this section show how much greenwashing content a person
might have seen on Facebook and Instagram as a proportion of all climate
change-related content in the country.

To evaluate hypothesis 2, we fit linear regression models of our
impression-weighted greenwashing scores as an outcome along with
covariates proposed by the literature as being important for explaining
government policy towards climate change, including Varieties of
Democracy scores (V-DEM). We include in our models the V-DEM executive
constraints index and polyarchy index, average CO\(_2\) consumption per
capita and oil and gas revenues per capita. Because this analysis is
exploratory, we fit a variety of models with and without other
covariates as controls and with and without interactions, which we
report in Table~\ref{tbl-models}.

We first note that we have countries present in our data across the full
range of the V-DEM polyarchy score with a minimum value of 0.016 (Saudi
Arabia) and a maximum value of 0.916 (Denmark). As such, we have
reasonable global coverage with which to test this hypothesis, even
though we acknowledge that the number of ads in non-Western countries is
considerably less than Western countries.

The results in Table~\ref{tbl-models} reveal that our hypothesis 2
relating to country-level regime type is largely falsified. We see
little evidence of any kind of systematic association between democratic
values or rule of law and greenwashing scores. We account for fuel
resources and carbon output as possible confounding variables, but we do
not see any difference in reported results when we do so, nor do we see
evidence of strong interaction effects. For these reasons, we believe
that institutional explanations---at least at the country level---do not
appear to be sufficient to explain this type of behavior.

The main caveat is that we do find a robust relationship between
\emph{lower} CO\(_2\) consumption and a higher ratio of greenwashing
content. This is a surprising finding as we might imagine that countries
with less CO\(_2\) output would also be less important targets for
greenwashing-related advertising. However, we note that our ad database
captures both ads with and without greenwashing, and there are a
substantial number of actors in countries like the United States that
are attempting to influence climate change discourse in a very different
direction than fossil fuel companies. As such, when measuring
impression-weighted average greenwashing scores by country, we find a
higher ratio of such content in places that perhaps we would not expect
initially.

\begin{table}

\caption{\label{tbl-models}Regression Models of Average Country
Greenwashing Scores}

\centering{

\centering
\begin{talltblr}[         
entry=none,label=none,
note{}={Table shows OLS regression coefficients (standard errors in parentheses) of country-level covariates on country-level average political greenwashing scores that have been weighted as a proportion of the total impressions for all climate-related ads within the given country. Results that are statistically significant at the $p$<0.05 level are denoted with an *.},
]                     
{                     
width={0.9\linewidth},
colspec={X[]X[]X[]X[]X[]X[]X[]X[]X[]},
cell{3}{1}={c=9}{},cell{9}{1}={c=9}{},
row{3,9}={}{font=\fontsize{0.7em}{1em}\selectfont,},
row{2}={}{halign=c, font=\fontsize{0.7em}{1em}\selectfont,},
cell{1}{1}={}{font=\fontsize{0.7em}{1em}\selectfont,},
cell{1}{3}={}{font=\fontsize{0.7em}{1em}\selectfont,},
cell{1}{4}={}{font=\fontsize{0.7em}{1em}\selectfont,},
cell{1}{5}={}{font=\fontsize{0.7em}{1em}\selectfont,},
cell{1}{6}={}{font=\fontsize{0.7em}{1em}\selectfont,},
cell{1}{7}={}{font=\fontsize{0.7em}{1em}\selectfont,},
cell{1}{8}={}{font=\fontsize{0.7em}{1em}\selectfont,},
cell{1}{9}={}{font=\fontsize{0.7em}{1em}\selectfont,},
cell{4}{2}={}{font=\fontsize{0.7em}{1em}\selectfont,},
cell{4}{3}={}{font=\fontsize{0.7em}{1em}\selectfont,},
cell{4}{4}={}{font=\fontsize{0.7em}{1em}\selectfont,},
cell{4}{5}={}{font=\fontsize{0.7em}{1em}\selectfont,},
cell{4}{6}={}{font=\fontsize{0.7em}{1em}\selectfont,},
cell{4}{7}={}{font=\fontsize{0.7em}{1em}\selectfont,},
cell{4}{8}={}{font=\fontsize{0.7em}{1em}\selectfont,},
cell{4}{9}={}{font=\fontsize{0.7em}{1em}\selectfont,},
cell{5}{2}={}{font=\fontsize{0.7em}{1em}\selectfont,},
cell{5}{3}={}{font=\fontsize{0.7em}{1em}\selectfont,},
cell{5}{4}={}{font=\fontsize{0.7em}{1em}\selectfont,},
cell{5}{5}={}{font=\fontsize{0.7em}{1em}\selectfont,},
cell{5}{6}={}{font=\fontsize{0.7em}{1em}\selectfont,},
cell{5}{7}={}{font=\fontsize{0.7em}{1em}\selectfont,},
cell{5}{8}={}{font=\fontsize{0.7em}{1em}\selectfont,},
cell{5}{9}={}{font=\fontsize{0.7em}{1em}\selectfont,},
cell{6}{2}={}{font=\fontsize{0.7em}{1em}\selectfont,},
cell{6}{3}={}{font=\fontsize{0.7em}{1em}\selectfont,},
cell{6}{4}={}{font=\fontsize{0.7em}{1em}\selectfont,},
cell{6}{5}={}{font=\fontsize{0.7em}{1em}\selectfont,},
cell{6}{6}={}{font=\fontsize{0.7em}{1em}\selectfont,},
cell{6}{7}={}{font=\fontsize{0.7em}{1em}\selectfont,},
cell{6}{8}={}{font=\fontsize{0.7em}{1em}\selectfont,},
cell{6}{9}={}{font=\fontsize{0.7em}{1em}\selectfont,},
cell{7}{2}={}{font=\fontsize{0.7em}{1em}\selectfont,},
cell{7}{3}={}{font=\fontsize{0.7em}{1em}\selectfont,},
cell{7}{4}={}{font=\fontsize{0.7em}{1em}\selectfont,},
cell{7}{5}={}{font=\fontsize{0.7em}{1em}\selectfont,},
cell{7}{6}={}{font=\fontsize{0.7em}{1em}\selectfont,},
cell{7}{7}={}{font=\fontsize{0.7em}{1em}\selectfont,},
cell{7}{8}={}{font=\fontsize{0.7em}{1em}\selectfont,},
cell{7}{9}={}{font=\fontsize{0.7em}{1em}\selectfont,},
cell{8}{2}={}{font=\fontsize{0.7em}{1em}\selectfont,},
cell{8}{3}={}{font=\fontsize{0.7em}{1em}\selectfont,},
cell{8}{4}={}{font=\fontsize{0.7em}{1em}\selectfont,},
cell{8}{5}={}{font=\fontsize{0.7em}{1em}\selectfont,},
cell{8}{6}={}{font=\fontsize{0.7em}{1em}\selectfont,},
cell{8}{7}={}{font=\fontsize{0.7em}{1em}\selectfont,},
cell{8}{8}={}{font=\fontsize{0.7em}{1em}\selectfont,},
cell{8}{9}={}{font=\fontsize{0.7em}{1em}\selectfont,},
cell{10}{2}={}{font=\fontsize{0.7em}{1em}\selectfont,},
cell{10}{3}={}{font=\fontsize{0.7em}{1em}\selectfont,},
cell{10}{4}={}{font=\fontsize{0.7em}{1em}\selectfont,},
cell{10}{5}={}{font=\fontsize{0.7em}{1em}\selectfont,},
cell{10}{6}={}{font=\fontsize{0.7em}{1em}\selectfont,},
cell{10}{7}={}{font=\fontsize{0.7em}{1em}\selectfont,},
cell{10}{8}={}{font=\fontsize{0.7em}{1em}\selectfont,},
cell{10}{9}={}{font=\fontsize{0.7em}{1em}\selectfont,},
cell{11}{2}={}{font=\fontsize{0.7em}{1em}\selectfont,},
cell{11}{3}={}{font=\fontsize{0.7em}{1em}\selectfont,},
cell{11}{4}={}{font=\fontsize{0.7em}{1em}\selectfont,},
cell{11}{5}={}{font=\fontsize{0.7em}{1em}\selectfont,},
cell{11}{6}={}{font=\fontsize{0.7em}{1em}\selectfont,},
cell{11}{7}={}{font=\fontsize{0.7em}{1em}\selectfont,},
cell{11}{8}={}{font=\fontsize{0.7em}{1em}\selectfont,},
cell{11}{9}={}{font=\fontsize{0.7em}{1em}\selectfont,},
cell{12}{2}={}{font=\fontsize{0.7em}{1em}\selectfont,},
cell{12}{3}={}{font=\fontsize{0.7em}{1em}\selectfont,},
cell{12}{4}={}{font=\fontsize{0.7em}{1em}\selectfont,},
cell{12}{5}={}{font=\fontsize{0.7em}{1em}\selectfont,},
cell{12}{6}={}{font=\fontsize{0.7em}{1em}\selectfont,},
cell{12}{7}={}{font=\fontsize{0.7em}{1em}\selectfont,},
cell{12}{8}={}{font=\fontsize{0.7em}{1em}\selectfont,},
cell{12}{9}={}{font=\fontsize{0.7em}{1em}\selectfont,},
cell{13}{2}={}{font=\fontsize{0.7em}{1em}\selectfont,},
cell{13}{3}={}{font=\fontsize{0.7em}{1em}\selectfont,},
cell{13}{4}={}{font=\fontsize{0.7em}{1em}\selectfont,},
cell{13}{5}={}{font=\fontsize{0.7em}{1em}\selectfont,},
cell{13}{6}={}{font=\fontsize{0.7em}{1em}\selectfont,},
cell{13}{7}={}{font=\fontsize{0.7em}{1em}\selectfont,},
cell{13}{8}={}{font=\fontsize{0.7em}{1em}\selectfont,},
cell{13}{9}={}{font=\fontsize{0.7em}{1em}\selectfont,},
cell{4}{1}={}{preto={\hspace{1em}}, font=\fontsize{0.7em}{1em}\selectfont,},
cell{5}{1}={}{preto={\hspace{1em}}, font=\fontsize{0.7em}{1em}\selectfont,},
cell{6}{1}={}{preto={\hspace{1em}}, font=\fontsize{0.7em}{1em}\selectfont,},
cell{7}{1}={}{preto={\hspace{1em}}, font=\fontsize{0.7em}{1em}\selectfont,},
cell{8}{1}={}{preto={\hspace{1em}}, font=\fontsize{0.7em}{1em}\selectfont,},
cell{10}{1}={}{preto={\hspace{1em}}, font=\fontsize{0.7em}{1em}\selectfont,},
cell{11}{1}={}{preto={\hspace{1em}}, font=\fontsize{0.7em}{1em}\selectfont,},
cell{12}{1}={}{preto={\hspace{1em}}, font=\fontsize{0.7em}{1em}\selectfont,},
cell{13}{1}={}{preto={\hspace{1em}}, font=\fontsize{0.7em}{1em}\selectfont,},
cell{1}{2}={c=8,}{font=\fontsize{0.7em}{1em}\selectfont, halign=c,},
hline{4}={1,2,3,4,5,6,7,8,9}{solid, lightgray, 0.1em},
hline{10}={1,2,3,4,5,6,7,8,9}{solid, lightgray, 0.1em},
}                     
\toprule
& Model &  &  &  &  &  &  &  \\ \cmidrule[lr]{2-9}
Results & (1) & (2) & (3) & (4) & (5) & (6) & (7) & (8) \\ \midrule 
Variables &&&&&&&& \\
Intercept & -0.31 (0.25) & -0.57 (0.17) & -0.75 (0.25) & -0.67 (0.16) & -0.48 (0.19) & -0.7 (0.15) & -0.27 (0.19) & -0.62 (0.16) \\
Exec. Constraints & -0.35 (0.4) & -0.46 (0.25) & - & - & - & - & -0.43 (0.26) & -0.37 (0.23) \\
Polyarchy & - & - & 0.54 (0.5) & -0.36 (0.28) & -0.1 (0.3) & -0.28 (0.27) & - & - \\
CO₂ pc & -0.03 (0.02) & - & -0.01 (0.02) & - & -0.04 (0.01) & - & -0.04 (0.01) & - \\
Oil pc & - & -37.29 (40.32) & - & -28.3 (39.5) & - & 1.8 (10.16) & - & 1.71 (10.06) \\
Model Fit &&&&&&&& \\
R² & 0.11 & 0.02 & 0.11 & 0.01 & 0.09 & 0.01 & 0.11 & 0.02 \\
Adj. R² & 0.08 & 0 & 0.08 & -0.01 & 0.07 & 0 & 0.09 & 0 \\
DF & 107 & 156 & 107 & 156 & 108 & 157 & 108 & 157 \\
F-Stat & 4.35 & 1.22 & 4.31 & 0.62 & 5.16 & 0.61 & 6.55 & 1.33 \\
\bottomrule
\end{talltblr}

}

\end{table}%

In SI section 2 we also report average levels of impression-weighted
political greenwashing separately for countries, major cities, U.S.
states, and age groups. There is substantial heterogeneity across
regions, and we also show that targeted ads occur at higher levels for
older demographics.

\subsubsection{Hypothesis 3: Sub-national Covariates and Greenwashing in
the United
States}\label{hypothesis-3-sub-national-covariates-and-greenwashing-in-the-united-states}

Similar to hypothesis 2, we fit linear models of impression-weighted
greenwashing scores for ads targeted toward individual US counties,
examining associations with covariates that explain local attitudes
toward climate change. Among our covariates is an aggregated measure of
community exposure and sensitivity to the transition away from fossil
fuels, developed by Raimi, Carley, and Konisky (2022). The authors use a
number of metrics (e.g., geological, climate, socioeconomic,
educational) to develop measures identifying the communities most likely
to be affected and harmed by the transition toward clean energy sources.
Specifically, we rely on an aggregate score combining communities'
sensitivity and exposure to the energy transition across coal, oil, and
natural gas. The scores, developed for census tracts, were then
aggregated to the county-level based on a population-weighted average.
We believe these \emph{transition risk scores} provide a reasonable
measurement of the distribution of costs associated with a transition to
renewable energy. In addition to these scores, we include county
partisanship, measured using the proportion of votes won by Republicans
in the 2024 election as a share of total Democratic and Republican votes
(2018); 2023 county gross domestic product (GDP), obtained from the
Bureau of Economic Analysis; and 2024 county population.\footnote{The
  Bureau of Economic Analysis's most recently published county gross
  domestic product figures in 2023. Population data was obtained from
  the US Census. Using 2020 presidential vote shares or 2023 population
  does not meaningfully change our results.}

Because this modeling is exploratory--we do not have a clean causal
identification strategy such as randomization of greenwashing scores--we
do our best to avoid well-known pitfalls of applied regression modeling
such as post-treatment bias (Montgomery, Nyhan, and Torres 2018) and
induced collider bias (Hünermund and Louw 2020) that can come from the
over-use of control variables. For these reasons, we fit models both
with and without adjustment variables both singly and jointly so that we
can test for obscure causal relationships between variables in the
model. While this does not in any way prove causal identification, it
can help uncover sensitivity of our results to a specific set of
adjustment variables. Regardless, the results in this section should be
interpreted as exploratory associations, albeit at a much lower level of
granularity than our country-level results.

\begin{table}

\caption{\label{tbl-gw-regressions-1}Regression Models of US County
Greenwashing Scores: Main Effects}

\centering{

 \centering
  \small
\begin{tabular}{@{\extracolsep{5pt}}lcccc}
\\[-1.8ex]\hline
\hline \\[-1.8ex]
 & \multicolumn{4}{c}{\textit{Dependent variable:}} \\
\cline{2-5}
\\[-1.8ex] & \multicolumn{4}{c}{Greenwashing Score} \\
\\[-1.8ex] & (1) & (2) & (3) & (4)\\
\hline \\[-1.8ex]
Transition & 0.3$^{***}$ & & & \\
 & (0.1) & & & \\
Republican & & 0.8$^{***}$ & & \\
 & & (0.2) & & \\
GDP & & & -0.0 & \\
 & & & (0.0) & \\
Population & & & & -0.0 \\
 & & & & (0.000) \\
Constant & -0.3$^{***}$ & -0.7$^{***}$ & -0.3$^{***}$ & -0.3$^{***}$ \\
 & (0.04) & (0.04) & (0.04) & (0.04) \\
\hline \\[-1.8ex]
Observations & 2,253 & 2,247 & 2,230 & 2,247 \\
R$^{2}$ & 0.01 & 0.01 & 0.000 & 0.000 \\
Adjusted R$^{2}$ & 0.000 & 0.000 & -0.000 & -0.000 \\
Residual Std. Error & 1.7 (df = 2251) & 1.7 (df = 2245) & 1.7 (df = 2228) & 1.7 (df = 2245) \\
F Statistic & 12.8$^{***}$ (df = 1; 2251) & 30.4$^{***}$ (df = 1; 2245) & 0.003 (df = 1; 2228) & 0.02 (df = 1; 2245) \\
\hline
\hline \\[-1.8ex]
\textit{Note:} & \multicolumn{4}{r}{$^{*}$p$<$0.1; $^{**}$p$<$0.05; $^{***}$p$<$0.01} \\
\end{tabular}

}

\end{table}%

\begin{table}

\caption{\label{tbl-gw-regressions-2}Regression Models of US County
Greenwashing Scores: Interaction Effects}

\centering{

 \centering
\small
\begin{tabular}{@{\extracolsep{5pt}}lcccc}
\\[-1.8ex]\hline
\hline \\[-1.8ex]
 & \multicolumn{3}{c}{\textit{Dependent variable:}} \\
\cline{2-4}
\\[-1.8ex] & \multicolumn{3}{c}{Greenwashing Score} \\
\\[-1.8ex] & (1) & (2) & (3)\\
\hline \\[-1.8ex]
Transition & 0.335$^{***}$ & 0.674$^{***}$ & 0.658$^{***}$ \\
 & (0.070) & (0.166) & (0.169) \\
GDP & & & -$-$0.000 \\
 & & & (0.000) \\
Republican & $-$3.819$^{***}$ & $-$3.671$^{***}$ & $-$3.969$^{***}$ \\
 & (0.719) & (0.721) & (0.728) \\
Republican$^2$ & 4.844$^{***}$ & 4.920$^{***}$ & 5.255$^{***}$ \\
 & (0.719) & (0.719) & (0.729) \\
GDP$^2$ & & & 0.000 \\
 & & & (0.000) \\
Population & & & $-$0.0000 \\
 & & & (0.00000) \\
Population$^2$ & & & $-$0.000 \\
 & & & (0.000) \\
Transition:Republican & & $-$0.683$^{**}$ & $-$0.656$^{**}$ \\
& & (0.304) & (0.307) \\
Constant & $-$5.078 & $-$5.069 & \\
 & (0.160) & (0.167) & (0.168) \\
\hline \\[-1.8ex]
Observations & 2,247 & 2,247 & 2,218 \\
R$^{2}$ & 0.041 & 0.043 & 0.044 \\
Adjusted R$^{2}$ & 0.040 & 0.042 & 0.043 \\
Residual Std. Error & 1.630 (df = 2243) & 1.628 (df = 2242) & 1.630 (df = 2209) \\
F Statistic & 32.033$^{***}$ (df = 3; 2243) & 25.333$^{***}$ (df = 4; 2242) & 13.380$^{***}$ (df = 8; 2209) \\
\hline
\hline \\[-1.8ex]
\textit{Note:} & \multicolumn{3}{r}{$^{*}$p$<$0.1; $^{**}$p$<$0.05; $^{***}$p$<$0.01} \\
\end{tabular}

}

\end{table}%

\begin{figure}

\caption{\label{fig-regressions-transition}Association Between
Transition Risk and Greenwashing Score}

\centering{

\pandocbounded{\includegraphics[keepaspectratio]{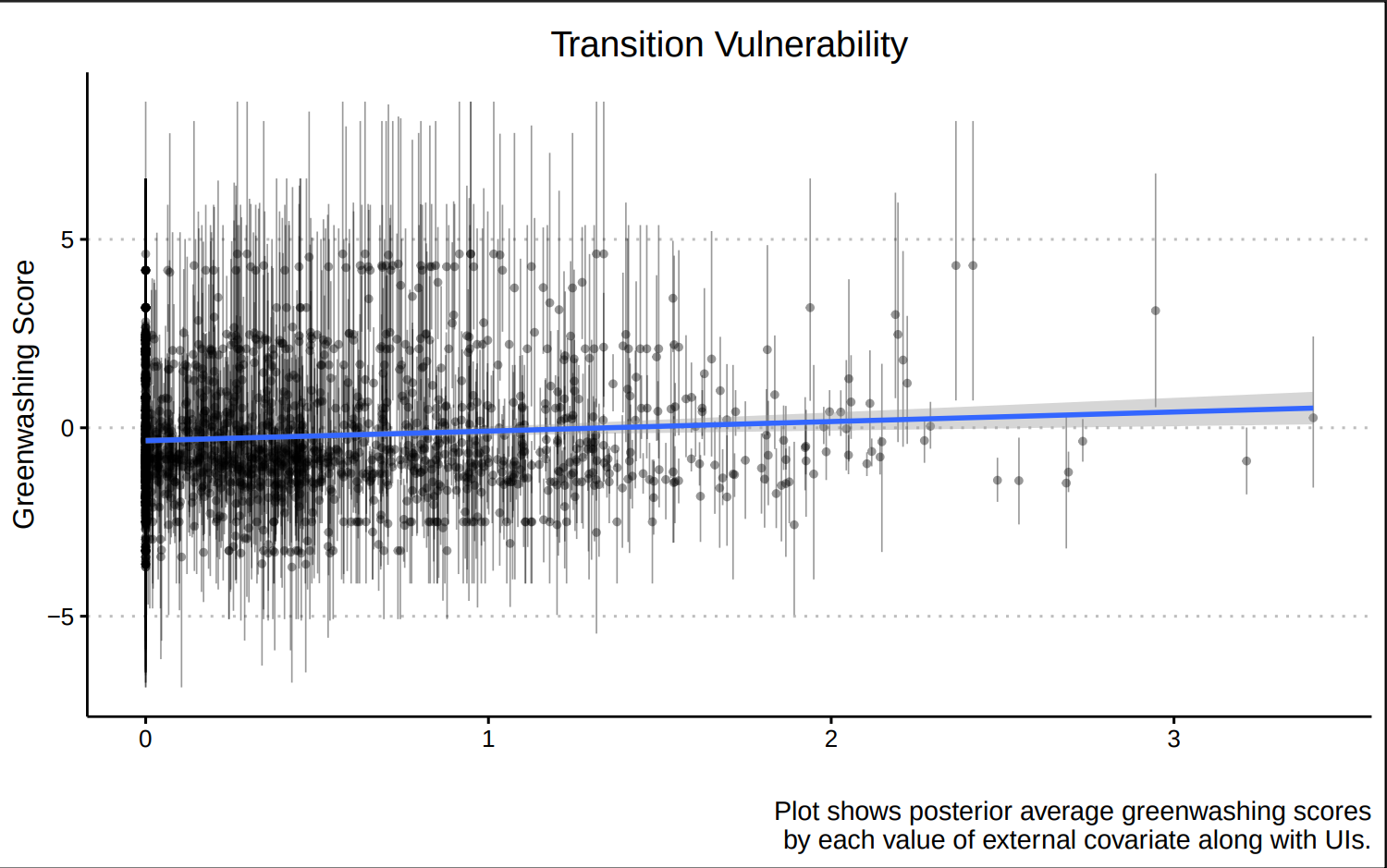}}

}

\end{figure}

Tables Table~\ref{tbl-gw-regressions-1} and
Table~\ref{tbl-gw-regressions-2} summarize our results, and
Figure~\ref{fig-regressions-transition} shows the distribution of a
random sample of 700 counties based on their transition risk and
greenwashing score. Concerning our hypothesis, there is a positive
association between transition risk and greenwashing scores across all
specifications. Climate-related advertisements targeted toward consumers
in counties vulnerable to the energy transition are more likely to
include greenwashing, even controlling for the county's population,
partisan composition, and GDP. Though we cannot exclude alternative
explanations, two seem more striking than others. First, fossil fuel
companies face strong incentives to exert local political pressure in
counties facing transition risk. As owners of local land and mineral
rights, companies are subject to municipal laws and regulations,
particularly if they aim to expand or upgrade existing facilities.
Homeowners may organize to block the construction of facilities near
their neighborhoods, and companies may face litigation from local
residents or workers affected by their operations. Second, since
residents in counties vulnerable to transition risk hold concentrated
interests in fossil fuel activity, companies may target them to ``shore
up'\,' a base of voters to oppose climate legislation.

\begin{figure}

\caption{\label{fig-effect-size}Marginal Effect Plot between Transition
Risk and 2024 Republican Vote Share}

\centering{

\pandocbounded{\includegraphics[keepaspectratio]{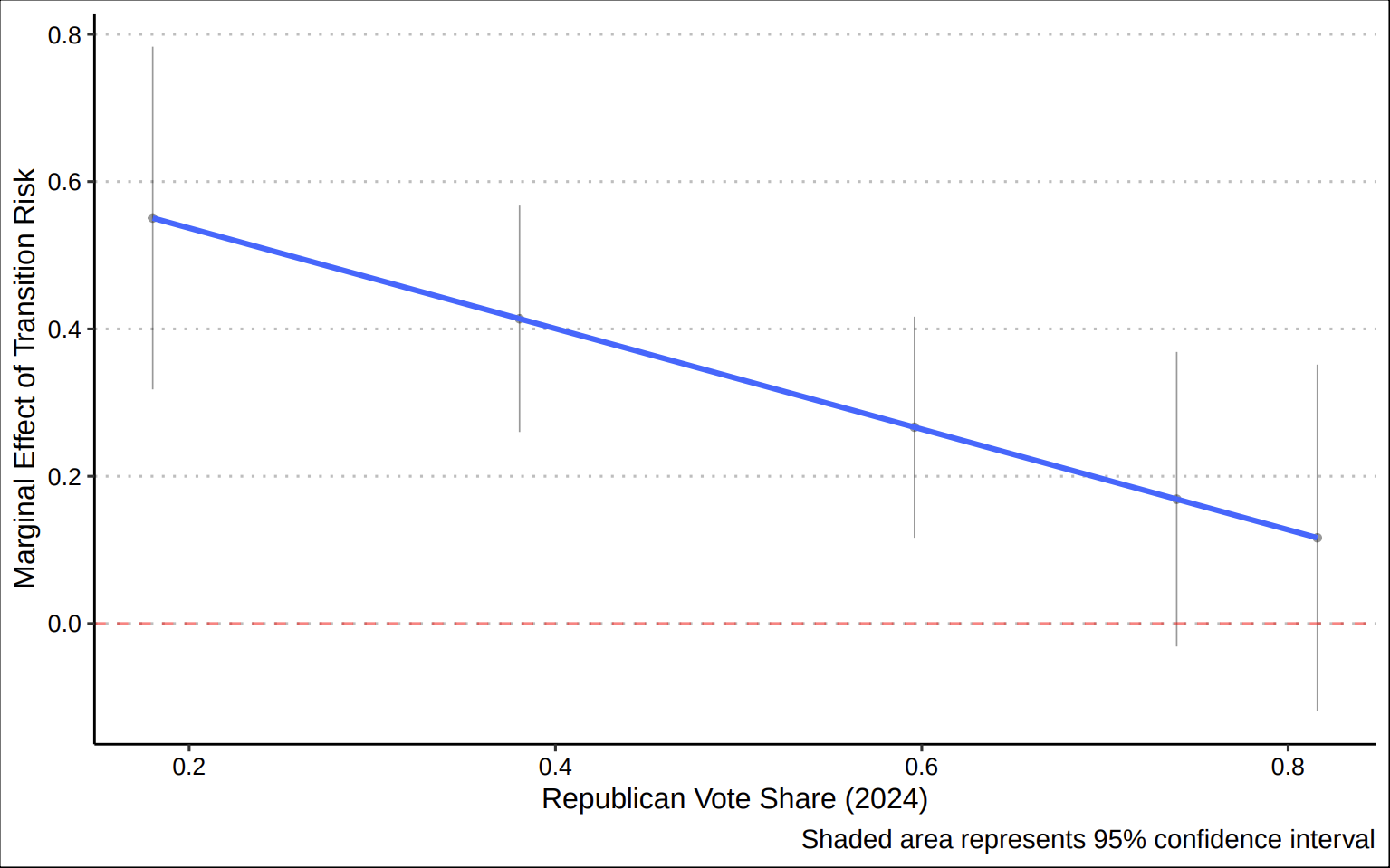}}

}

\end{figure}

Figure~\ref{fig-effect-size} illustrates changes in the marginal effect
of transition risk at different levels of partisanship. An increase in
transition risks is associated with sharper increases in greenwashing
scores in counties where smaller shares of voters supported Trump. Due
to their economic interests, voters in these counties may be more prone
to oppose climate legislation than those in other Democratic counties.
Equivalently, compared to other voters in counties vulnerable to
transition risk, these voters may be more susceptible to Democratic
messages supporting climate legislation. From either perspective, there
are strong incentives for emitting companies to persuade these voters
that their products and practices are less harmful than they appear.

\section{Conclusion}\label{conclusion}

In this paper we are able to show using granular data how fossil fuel
companies are targeting people directly in areas in which they may fear
adverse regulatory decisions. Contrary to our initial expectations, we
find these ads to be micro-targeted at places where oil companies have
fixed assets. We do not find much evidence for the idea that
greenwashing may be more prominent in more democratic countries,
although we do find evidence that greenwashing is more common among
older demographic groups who tend to vote at higher rates. We believe
that our measurement approach can unlock further insights into political
greenwashing as well as provide a signal detection method for quickly
and accurately classifying ads as potentially containing greenwashing.

\newpage

\section{References}\label{references}

\phantomsection\label{refs}

\end{document}